\documentclass[sigconf,natbib=true,anonymous=false]{acmart}
 \usepackage{enumitem}
  \usepackage{subcaption}
  \usepackage[export]{adjustbox}
  \usepackage{makecell}
  \usepackage{multirow}
  \usepackage{lineno}
  \usepackage{bm}
\AtBeginDocument{%
  \providecommand\BibTeX{{%
    \normalfont B\kern-0.5em{\scshape i\kern-0.25em b}\kern-0.8em\TeX}}}

\setcopyright{acmcopyright}
\copyrightyear{2022}
\acmYear{2022}
\acmDOI{0.0/0.0}

\begin{document}

\title{Sparse Attentive Memory Network for Click-through Rate Prediction with Long Sequences}


\author{Qianying Lin}
\affiliation{%
  \institution{Alibaba Group}
  \streetaddress{Wenyi West Road 969 Hao, Yuhang District}
  \city{Hangzhou}
  \country{China}}
\email{qianying.lqy@alibaba-inc.com}

\author{Wen-Ji Zhou}
\affiliation{%
  \institution{Alibaba Group}
  \city{Hangzhou}
  \country{China}}
\email{eric.zwj@alibaba-inc.com}

\author{Yanshi Wang}
\affiliation{%
  \institution{Alibaba Group}
  \city{Hangzhou}
  \country{China}}
\email{yanshi.wys@alibaba-inc.com}

\author{Qing Da}
\affiliation{%
  \institution{Alibaba Group}
  \city{Hangzhou}
  \country{China}}
\email{daqing.dq@alibaba-inc.com}

\author{Qing-Guo Chen}
\affiliation{%
  \institution{Alibaba Group}
  \city{Hangzhou}
  \country{China}}
\email{qingguo.cqg@alibaba-inc.com}

\author{Bing Wang}
\affiliation{%
  \institution{Alibaba Group}
  \city{Hangzhou}
  \country{China}}
\email{lingfeng.wb@alibaba-inc.com}

\renewcommand{\shortauthors}{Lin, et al.}
\begin{abstract}
Sequential recommendation predicts users' next behaviors with their historical interactions. Recommending with longer sequences improves recommendation accuracy and increases the degree of personalization. As sequences get longer, existing works have not yet addressed the following two main challenges. Firstly, modeling long-range intra-sequence dependency is difficult with increasing sequence lengths. Secondly, it requires efficient memory and computational speeds. In this paper, we propose a Sparse Attentive Memory (SAM) network for long sequential user behavior modeling. SAM supports efficient training and real-time inference for user behavior sequences with lengths on the scale of thousands. In SAM, we model the target item as the query and the long sequence as the knowledge database, where the former continuously elicits relevant information from the latter. SAM simultaneously models target-sequence dependencies and long-range intra-sequence dependencies with $O(L)$ complexity and $O(1)$ number of sequential updates, which can only be achieved by the self-attention mechanism with $O(L^2)$ complexity. Extensive empirical results demonstrate that our proposed solution is effective not only in long user behavior modeling but also on short sequences modeling. Implemented on sequences of length 1000, SAM is successfully deployed on one of the largest international E-commerce platforms. This inference time is within 30ms, with a substantial 7.30\% click-through rate improvement for the online A/B test. To the best of our knowledge, it is the first end-to-end long user sequence modeling framework that models intra-sequence and target-sequence dependencies with the aforementioned degree of efficiency and successfully deployed on a large-scale real-time industrial recommender system.
\end{abstract}



\maketitle

\section{Introduction}
Click-through rate (CTR) prediction is a core task in recommender systems. User sequential modeling is the key to mine users' interest for accurate predictions. The sequences used are usually truncated to users' most recent 50 to 100 behaviors \citep{dien, din}. As user behavior records accumulate, it is possible to model longer user sequences. The introduction of long-term interests improves both recommendation accuracy and the degree of personalization. Yet as sequences get longer, particularly with lengths longer than 1000, the prediction task requires extraordinary long-range dependency modeling, efficient memory, acceptable training speed and real-time inference.\\
\citet{hidasi} employ Recurrent Neural Networks (RNNs) for sequential recommenders, summarizing previous actions with a hidden state for the next action prediction. The long short-term memory (LSTM) is a special class in RNNs that models sequential behaviors \citep{Hochreiter}. \citet{Graves} prove that LSTM forgets quickly and fails to generalize to sequences longer than 20. Many empirical results also verify that RNN-based sequential recommenders do not perform as well as attention-based methods since the hidden state forgets long-term information quickly \citep{atrank, kang, din,MIMN}. \\
Lately, the self-attention mechanism has proven to benefit a wide range of application domains, such as machine translation \citep{transformer}, speech recognition \citep{chan}, reading comprehension \citep{cui,lin} and computer vision \citep{xu,parmar}. The self-attention mechanism attends to different positions in the sequence, captures the most important features and allows the model to handle long-range intra-sequence dependencies. Self-Attentive Sequential Recommendation (SASRec) adapts the self-attentive Transformer architecture for sequential recommenders and outperforms convolution-based and recurrence-based methods empirically \citep{kang}. \\
Two problems arise applying SASRec to long sequential recommender systems. Firstly, the memory complexity and the computational complexity are both quadratic with respect to the sequence length. The quadratic computational complexity might not be the major bottleneck since the self-attention mechanism allows for parallelization. Yet, the $O(L^2)$ memory complexity makes it infeasible to handle long sequences. Research on efficient self-attention is based on either sparse attention \citep{sac, informer, child, reformer} or approximated attention \citep{linformer, qin2022cosformer}, and consequently incompetent against the original Transformer. Furthermore, these methods are experimented in Natural Language Processing (NLP) or Computer Vision (CV), with no proven effective adaptations on recommender systems. Secondly, the self-attention mechanism is performed in a fixed fully-connected structure, which can be non-optimal for the click-through rate prediction task. SASRec encodes user sequences with Transformer and does not involve the target item for encoding. \\
Deep Interest Network (DIN) is designed to model user sequential behaviors \citep{din}. It adaptively learns the user interest representation from historical behaviors with respect to a particular target item. The space and time complexities for DIN are linear, but DIN cannot model intra-sequence dependencies. \\
Works succeeding DIN employ more complicated encoding mechanisms, which mostly rely on sequential updates. DIEN and MIMN perform sequential updates per incoming item, which imposes great difficulty on training and online serving\citep{dien,MIMN}.  \\
In this paper, we propose the Sparse Attentive Memory (SAM) network for long sequential user behavior modeling \footnote{The source codes are available at https://github.com/waldenlqy/SAM.}. In SAM, the target item acts as the query and the long sequence acts as the knowledge database, where the former continuously elicits relevant information from the latter. The contributions of this paper are summarized as follows:
\begin{itemize}[leftmargin=*]
\setlength{\itemsep}{0pt}
\setlength{\parskip}{0pt}
\item We propose the Sparse Attentive Memory (SAM) network, an end-to-end differentiable framework for long user sequential behavior modeling. It supports efficient training and real-time inference for user sequences with lengths on the scale of thousands. 
\item We derive a sparse attentive memory network to simultaneously model target-sequence dependencies and long-range intra-sequence dependencies with $O(L)$ complexity and $O(1)$ number of sequential updates. To the best of our knowledge, it is the first design to model intra-sequence and target-sequence dependencies with the aforementioned degree of efficiency. 
\item With greater computational and memory efficiency, SAM is deployed successfully on one of the largest international E-commerce platforms, with the number of items on the scale of hundreds of millions. Implemented on user sequences with length 1000 and deployed on GPU clusters, it supports real-time inference within 30ms. There is a significant 7.30\% CTR improvement over the DIN-based industrial baseline. 
\item Extensive experiments on both public benchmarks and the industrial dataset demonstrate our proposed solution's effectiveness not limited to long user behavior modeling but also on short sequences modeling.  

\end{itemize}


\section{Related Work}
\textbf{Sequential Recommender Systems}. Sequential recommender systems predict the user's next clicking behavior based on his past activities. Recurrent Neural Networks (RNNs) are introduced for sequential recommenders \citep{wu,hidasi}. Due to their sequential nature, RNN-based methods are difficult to parallelize. RNNs also suffer from the problem of fast forgetting \citep{Graves}. Attention is first introduced in the encoder-decoder framework, for better positional alignments in the machine translation task \citep{bahdanau}. Researchers prove empirically that the self-attention mechanism with timestamp encodings can replace RNNs to encode sequences, with significantly less training time \citep{transformer}. Attention-based sequential models proliferate in many other tasks, such as computer vision \citep{xu}, reading comprehension \citep{cui,lin} and speech recognition \citep{chan}. Attention-based recommender systems include methods based on self-attention \citep{kang, atrank}, methods based on target attention\citep{din} and the integration between recurrence-based methods and attention-based methods \citep{dien}. \\

\noindent \textbf{Memory Networks}. Memory Networks have wide applications in Question-Answering (QA) for NLP tasks, finding facts related to a particular query from a knowledge database \citep{chaudhari}. It can be viewed as a generalization to the attention mechanism with an external memory component. Neural Turing Machines (NTM) introduces the addressing-read-write mechanism for the memory search and update process \citep{Graves}. \cite{weston} proposes the general architecture for Memory Networks. DMN, DMTN and DMN+ are the subsequent research \citep{kumar, xiong,sachithanandam}. In recommender systems, MIMN utilizes the NTM architecture and uses GRU-based controllers to update user memory slots with each new clicked item \citep{MIMN}. SAM is different from the above architectures. Though we also keep an external memory vector, we do not update the memory with each new input, hence the number of sequential operations is $O(1)$.



\section{Problem Formulation}
The recommender system models the user-item interaction as a matrix $C = \{c_{mn}\}_{M \times N}$, where $M$ and $N$ are the total number of users and items respectively. The interaction is either explicit ratings \citep{Koren} or implicit feedback \citep{agarwal}. The click-through rate prediction task is usually based on implicit feedback. We denote $u \in U$ as user and $i \in I$ as item, and the user $u_m$ clicking on the item $i_n$ makes $c_{mn}$ 1 and others 0. User sequential modeling predicts the probability of a user $u\in U$ clicking on the target item $i \in I$ based on his past behaviors, $i_1, i_2, ..., i_{L}$, where $L$ is the length of the user sequence. The sequence is usually in chronological order. Sequential recommenders usually use the most recent 50 to 100 behaviors. Our paper focuses on long user sequences, where the length of the user behaviors $L$ is on the scale of thousands. 

\section{Sparse Attentive Memory Network}
\begin{figure*}
\centering
\includegraphics[scale=0.142]{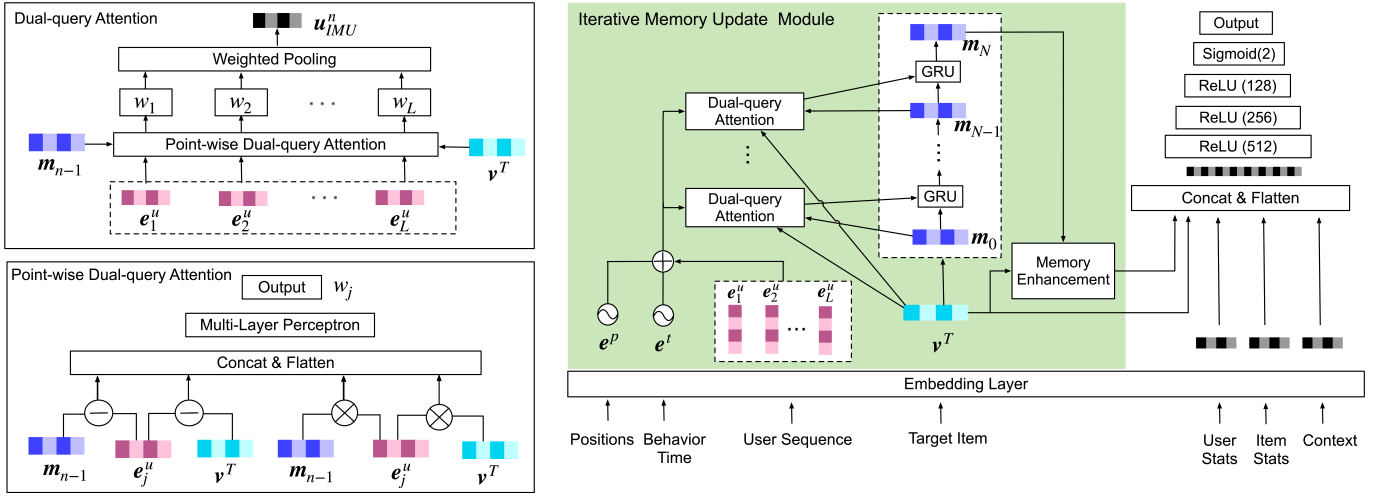}
\caption{Sparse Attentive Memory (SAM) network architecture overview. The Iterative Memory Update module (in green) is our main contribution. The details about the dual-query attention are on the left.
}
\label{fig:overview}
\end{figure*}



\begin{figure}
\centering
\includegraphics[width=\columnwidth]{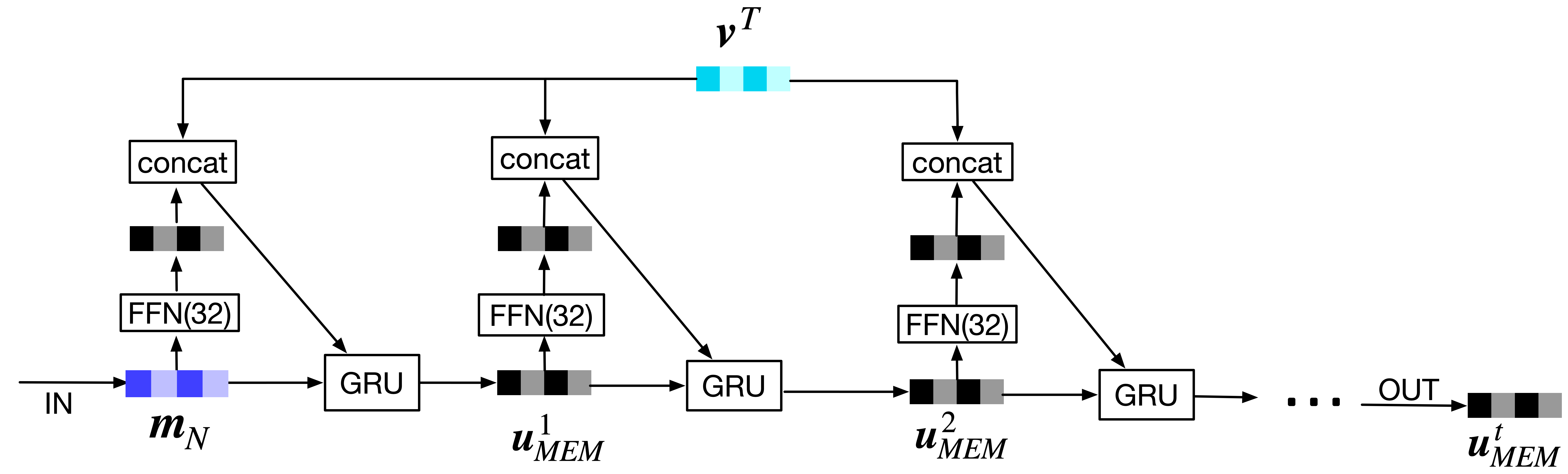}
\caption{Memory Enhancement module}
\label{fig:mem_enhance}
\end{figure}

 The industrial recommender system is usually a two-staged system, consisting of the retrieval stage and the rank stage. Compared to the retrieval task where it learns the probability to click each item from billions of candidates during training and performs an Approximate Nearest Neighbor (ANN) search during the inference stage, the rank task has access to the target item to be scored. In view of this, SAM frames the rank task as a Question-Answering (QA) task, where the target item resembles the question and the long sequence resembles the knowledge base. The task is to find relevant facts about the target item from the knowledge base of long sequence. We illustrate the Sparse Attentive Memory (SAM) network architecture in Fig.\ref{fig:overview} and discuss the framework as follows. \\

\subsection{Encoder Layer}
The user behavior sequence can be split into three parts, the clicked item sequence, the timestamp sequence and the positional sequence. For the clicked item sequence $\boldsymbol{e}(B^i)=\{\boldsymbol{e}^i_1,\boldsymbol{e}^i_2,...,\boldsymbol{e}^i_L\}$, $\boldsymbol{e}^i_j \in \mathbb{R}^{d_i}$ is obtained by concatenating the j-th clicked item feature embedding vectors, including item id, category id, shop id and brand id, i.e.,
$\boldsymbol{e}^i_j=[\boldsymbol{e}_{itemId}||\boldsymbol{e}_{cateId}||\boldsymbol{e}_{shopId}||\boldsymbol{e}_{brandId}]$, where $||$ is the vector concatenation operator and $d_i$ is the dimension for the concatenated embedding vector. Different users have different action patterns, thus the action time contains important temporal information. Since it is difficult to learn a good embedding directly with continuous time features, we bucketize the timestamp into multiple granularities and perform categorical feature look-ups. We slice the elapsed time with respect to the ranking time into intervals whose gap length increases exponentially. In other words, we map the time in range [0,1), [1,2), [2,4), ..., [$2^k$, $2^{k+1}$) to categorical features 0,1,2,...$k+1$ and perform categorical feature look-ups to obtain the absolute timestamp sequence  $\boldsymbol{e}(B^t)=\{\boldsymbol{e}^t_1,\boldsymbol{e}^t_2,...,\boldsymbol{e}^t_L\}$. Positional encodings are also added to represent the relative positions of sequence items. Since not all user sequences have length 1000, the positions are numbered in descending order of the serial number $\{L, L-1, ..., 1\}$ where the most recent behavior is always 1 to ensure the semantics are the same for the same recency of the behavior. The positional sequence $\boldsymbol{e}(B^p)=\{\boldsymbol{e}^p_1,\boldsymbol{e}^p_2,...,\boldsymbol{e}^p_L\}$is obtained by the categorical feature look-up on the numbered positions. The clicked item sequence, the timestamp sequence and the positional sequence are summed up on each position to obtain the final encoder layer representation $\boldsymbol{e}(B^u)=\{\boldsymbol{e}^u_1,\boldsymbol{e}^u_2,...,\boldsymbol{e}^u_L\}$, where the j-th user behavior $\boldsymbol{e}^u_j$ is obtained as $\boldsymbol{e}_{j}^{u} = \boldsymbol{e}_{j}^{i} \oplus \boldsymbol{e}_{j}^{t} \oplus \boldsymbol{e}_{j}^{p}$ and $\oplus$ denotes element-wise sum-up. The timestamp encoding $\boldsymbol{e}_{j}^{t} \in \mathbb{R}^{d_t}$ and the positional encoding $\boldsymbol{e}_{j}^{p} \in \mathbb{R}^{d_p}$ have the same dimension as that of the item embedding $\boldsymbol{e}_{j}^{i}$ to be directly summed. That is, $d_i=d_t=d_p$. The embedding vector for the target item $\boldsymbol{v}^T \in \mathbb{R}^{d_i}$ shares the id embedding look-up tables with the sequence item id embedding $\boldsymbol{e}_{j}^{i}$.

\subsection{Point-wise Dual-query Attention}
Long-range intra-sequence dependency modeling is important for long sequences. Since longer sequences contain more noises, the patterns within the long sequence are more difficult to mine. The self-attention mechanism is designed to capture intra-sequence dependencies. The canonical self-attention mechanism is in the form $Attention(Q,K,V)=softmax(\frac{QK^T}{\sqrt{d}})V$,where $Q,K,V$ are linear transformations of the input sequence. Yet the $O(L^2)$ space and time complexities make it not scalable to long sequences. The Point-wise Dual-query Attention (PDA) is the sparse attention mechanism that we propose to model intra-sequence dependencies in $O(L)$ space and time complexities. \\
As seen in Fig.\ref{fig:overview}, for the j-th user behavior item  $\boldsymbol{e}_{j}^{u} \in \mathbb{R}^{d_i}$, we apply an attention mechanism with both the target item $\boldsymbol{v}^T \in \mathbb{R}^{d_i}$ and the memory $\boldsymbol{m}_t \in \mathbb{R}^{d_i}$ as dual-queries to adaptively learn the weight for each behavior. We will introduce how the memory vector $\boldsymbol{m}_t \in \mathbb{R}^{d_i}$ is initialized and updated in the following section 4.3. \\
We define the feature vector $\boldsymbol{\alpha}_j \in \mathbb{R}^{d_\alpha}$ to capture the tripartite relations amongst the sequence item $\boldsymbol{e}_{j}^{u} \in \mathbb{R}^{d_i}$, the target item $\boldsymbol{v}^T \in \mathbb{R}^{d_i}$, and the memory embedding vector $\boldsymbol{m}_t \in \mathbb{R}^{d_i}$. 
\begin{equation}
 \boldsymbol{\alpha}_j (\boldsymbol{e}_{j}^{u}, \boldsymbol{v}^T, \boldsymbol{m}_t) = [\boldsymbol{e}_{j}^{u} \ominus \boldsymbol{m}_t \,||\, \boldsymbol{e}_{j}^{u} \ominus \boldsymbol{v}^T \,||\, \boldsymbol{e}_{j}^{u} \otimes \boldsymbol{m}_t \,||\, \boldsymbol{e}_{j}^{u} \otimes \boldsymbol{v}^T]
\end{equation}
where $\ominus$ denotes the element-wise subtraction operation and $\otimes$ denotes the element-wise multiplication operation. \\
We input each feature vector $\boldsymbol{\alpha}_j$ corresponding to the j-th behavior $\boldsymbol{e}_{j}^{u}$ into a two-layer point-wise feed-forward network. In other words, we employ the feed-forward attention operator with \textit{sigmoid} as the activation function on the input feature vector. 
\begin{equation}
 a_j(\boldsymbol{e}_{j}^{u}, \boldsymbol{v}^T, \boldsymbol{m}_t) = \sigma(\boldsymbol{W}^{(2)}\sigma(\boldsymbol{W}^{(1)} \boldsymbol{\alpha}_j (\boldsymbol{e}_{j}^{u}, \boldsymbol{v}^{T}, \boldsymbol{m}_t) + b^{(1)}) + b^{(2)}) 
\end{equation}
$\boldsymbol{W}^{(1)}\in\mathbb{R}^{d_i \times d_h}$, $\boldsymbol{W}^{(2)} \in \mathbb{R}^{d_h \times 1}$, $\boldsymbol{b}^{(1)}\in \mathbb{R}^{d_h}$ and $\boldsymbol{b}^{(2)} \in \mathbb{R}^{d_h}$ are learnable parameters shared across sequence items. $\sigma$ denotes the \textit{sigmoid} activation function. The fully-connected feed-forward network is applied to each sequence item separately and identically. We have also experimented with the \textit{softmax} activation function for the second layer. There is negligible change in model performance.\\
The dual-query attention uses both the target item and the memory vector as dual-queries to query the long sequence. As will be introduced in Section 4.3, the memory vector is updated with the retrieved sequence information. Querying the sequence with the memory vector models long-range intra-sequence dependencies. \\

\subsection{Iterative Memory Update Module}
While short-term memorization only requires limited memorization power, long-sequence memorization inevitably incurs the problem of gradually forgetting the early contents. Since the recurrent connection mechanism is limited in long-range dependency modeling \citep{sodhani2020toward}, additional architecture components are required to capture long-term user preferences. To this end, SAM maintains an external memory matrix $\boldsymbol{m}_t \in \mathbb{R}^{d_i}$ to expand the memorization capacity and memorize a user's long-term preferences. \\
Another challenge is how to design an effective memory update mechanism. Put mathematically, we need an effective abstraction function $f(.) : \mathbb{R}^{d_i+L} \rightarrow \mathbb{R}^{d_i}$ for the $n$-th memory update iteration as 
\begin{equation}
      \boldsymbol{m}_n \gets f( \boldsymbol{\cup} (\boldsymbol{e}(B^u)), \boldsymbol{m}_{n-1})
\end{equation}
where $\boldsymbol{\cup} (\boldsymbol{e}(B^u))$ is the set of user behavior sequence. Since long sequences contain much more information compared to short sequences, using fixed-size memory slots for memory abstraction inevitably leads to information loss. Maintaining a fixed-size first-in-first-out (FIFO) memory to cache the long-term information is reasonable in NLP tasks where related words are usually not far in the sentence \citep{rae2019compressive}, but in recommender systems the behavior sequence is not strictly ordered and users can exhibit seasonal periodic behaviors\citep{tan2021dynamic, yuan2019simple}. Therefore, instead of requiring the memory to \textit{memorize as much as possible}, we propose to give a clue so that the model can search for and memorize useful facts \textit{with the question}. Since research has empirically validated the importance of the target item in the rank task \citep{din,dien}, we propose to use the target item as the clue. We model the rank task as the Question-Answering (QA) task. The target item is the question and the long sequence is the knowledge base, with the task to find facts related to the question from the knowledge database. We introduce the memory abstraction and update process in detail as follows.\\
The initial memory $\boldsymbol{m}_0 \in \mathbb{R}^{d_i}$ is initialized from the target item vector $\boldsymbol{v}^T\in \mathbb{R}^{d_i}$, to model the stage where the question is presented and no sequence information has been included. \\
\begin{equation}
\boldsymbol{m}_0 = \boldsymbol{v}^T
\end{equation}
For iteration $n$, we apply weighted-sum pooling to the feature vector list of the user's sub behaviors to map it to the user representation vector $\boldsymbol{u}^n_{IMU}\in \mathbb{R}^{d_i}$. The weights for the weighted-sum pooling are derived from Eq.(2). 
\begin{equation}
\begin{aligned}
\boldsymbol{u}^n_{IMU} &= f(\boldsymbol{v}^T, \boldsymbol{\cup} (\boldsymbol{e}(B^u)),\boldsymbol{m}_{n-1}) \\&= \sum_{j=1}^{L} a_j(\boldsymbol{e}^{u}_{j}, \boldsymbol{v}^T, \boldsymbol{m}_{n-1})\boldsymbol{e}_{j}^{u} = \sum_{j=1}^{L}w_j\boldsymbol{e}_{j}^{u}
\end{aligned}
\end{equation} \\ 
Here we use weighted-sum pooling instead of sequential operations such as GRU and GRU with attentional update (AUGRU)\citep{dien} due to the following two reasons. Firstly, as aforementioned, the behavior sequence is not strictly ordered therefore we do not need sequential operations to model the strict order of sequence items. Secondly, though GRUs can also model intra-sequence dependencies, sequential updates hinder training and deployment for long sequences. The computational cost analysis in Section 5.6 validates the computational inefficiency with methods relying on sequential update operations. \\
Though we do not connect sequence items with  recurrent mechanisms, we use a Gated Recurrent Network (GRU) to model the memory update mechanism after each iteration. We choose GRU because we intend to use the update gate to adaptively determine what to forget and what to memorize. We abbreviate the computation for GRU as $\boldsymbol{h}_t=GRU(\boldsymbol{x}_t, \boldsymbol{h}_{t-1})$ where $\boldsymbol{h}_{t-1}$ is the vector representation for the last step and $\boldsymbol{x}_t$ is the input for the current step. \\ 
Each memory update takes place after a full pass of the sequence. We use the user interest representation vector $\boldsymbol{u}^n_{IMU}$ after iteration $n$ as the input to update the memory $\boldsymbol{m}_{n-1}$. 
\begin{equation}
      \boldsymbol{m}_n = GRU(\boldsymbol{u}^n_{IMU}, \boldsymbol{m}_{n-1})
\end{equation}
After $N$ iterations of the memory update mechanism, the final output from this module is $\boldsymbol{m}_N$. \\
A popular choice for industrial click-through rate prediction models is DIN\citep{din}. The target attention mechanism in DIN uses the target item to query sequence items to produce the weights for sequence item aggregation, therefore DIN only models target-sequence dependencies. On the contrary, SAM's memory vector $\boldsymbol{m}_{n}$ is updated with a weighted sum pooling of sequence items. Since the memory vector $\boldsymbol{m}_{n}$ contains information about the sequence items, querying sequence items with the memory vector models intra-sequence dependencies. \\
In other words, SAM models co-occurrence beyond the (target item $\boldsymbol{v}^T$, sequence item $\boldsymbol{e}^{u}_{j}$) pair. For example, the target item is rum and the user sequence contains lime and peppermint. With the target attention mechanism, both the attention weight between the pair (peppermint, rum) and that between the pair (lime, rum) are not high. In contrast, the memory vector in SAM is continuously updated with the weighted aggregation of sequence items, therefore it contains information about the peppermint and rum. When calculating the attention score for lime after the first memory update iteration, the memory of peppermint and rum awakens the item lime since the triplet (rum, peppermint, lime) is the recipe for Mojito and likely to co-occur multiple times. Hence, the likelihood to click rum increases with lime and peppermint in the sequence. While the target attention mechanism finds items that co-occur frequently with the target item, SAM finds the composite group of the user's behavior items for the user to click the target item.  \\

\subsection{Memory Enhancement Module}
The Memory Enhancement module takes the output from Iterative Memory Update module $\boldsymbol{m}_N$ as the input. It enhances the user memory $\boldsymbol{m}_N$ with the target item $\boldsymbol{v}^T$ repeatedly to elicit more clear memory specific to the target item and remove noises. \\
We use another GRU to model the memory enhancement process. The GRU's initial state is initialized from the memory after the Iterative Memory Update module, $\boldsymbol{u}^0_{MEM} = \boldsymbol{m}_N$. For each step, we apply a linear transformation $W^{u} \in \mathbb{R}^{d_i \times d_i}$ on the GRU's last hidden state $\boldsymbol{u}^{t-1}_{MEM}$, concatenate the transformed vector with the target item, and use the concatenated vector as the GRU's input.
\begin{equation}
      \boldsymbol{u}^{t}_{MEM} = GRU([\boldsymbol{W}^{u}\boldsymbol{u}^{t-1}_{MEM} \,||\, \boldsymbol{v}^T], \boldsymbol{u}^{t-1}_{MEM})
\end{equation}
where $||$ is the concatenation operator and $\boldsymbol{u}^t_{MEM}$ is the user representation after $t$ steps in the Memory Enhancement module. We illustrate the Memory Enhancement module in Fig.\ref{fig:mem_enhance}. \\
The final output $\boldsymbol{u}^{t}_{MEM}$ from the module is concatenated with the vector representations of other item and user features followed by a multilayer perceptron (MLP) encoder to produce the final logit. Sigmoid is applied on the logit to get the final prediction $\hat{y_i}$. We minimize the cross entropy loss function between the predicted $\hat{y_i}$ and the ground-truth $y_i$.


\section{Experiments}
This section presents the experimental setups, experimental results, ablation study, model analysis, computational cost and memory efficiency analysis, performance analysis on sequences of lengths up to 16K and hyper-parameter choices in detail. 

\subsection{Datasets and Experimental Setup}
\textbf{Amazon Dataset}. We collect two subsets from the Amazon product data, Books and Movies \citep{McAuley}. Books contains 295982 users, 647589 items and 6626872 samples. Movies contains 233282 users, 165851 items and 4829693 samples. We split each dataset into 80\% training and 20\% test data according to the behavior timestamp. The sequence embedding dimension is 16. The MLP layer size is $64 \times 32$. We use the Adam optimizer, with 0.001 learning rate \citep{KingmaB14}. The mini-batch size is 512. We use 2 parameter servers and 4 workers, with 10GiB memory for each worker. \\
\textbf{Industrial Dataset}. We collect traffic logs from a real-world E-commerce platform. The E-commerce platform has search and recommendation systems, with user click and purchase logs. We use 30-day samples for training and the samples of the following day for testing. With 0.1 sampling on negative samples, there are 1.68 billion training samples. The ratio of positive to negative samples is 1:2.24 in the training set. The test set contains 57 million data points. The id embedding dimension is 32. The hidden state dimensions for GRUs are 32. MLP layers are $512 \times 256 \times 128$. The mini-batch size is 512. We use the Adam optimizer, with 0.0001 as the learning rate. We use 5 parameter servers and 50 workers, with 75GiB memory for each worker. \\ \textbf{Evaluation Metric}. We use Area Under the Curve (AUC) to measure the model performance. For the CTR prediction task, it represents the probability that the model ranks a randomly chosen clicked instance higher than a randomly chosen unclicked instance. 

\subsection{Model Comparison}
While there is abundant research on click-through rate prediction, we select the relevant and representative baselines. Since our proposed method focuses on long sequence modeling, we do not include models on different topics such as xDeepFM and FFM which learn categorical feature interactions\citep{lian2018xdeepfm, juan2016field}. We exclude methods based on Graph Neural Networks (GNNs) since research has shown their computational complexity limits the scalability to longer sequences \citep{gnn,magnn,guo2020intention}. We also exclude models that integrate long-term and short-term interests since our method models long-term interests and adding short-term interests modeling with highly complicated methods results in unfair comparisons \citep{tan2021dynamic, yu2019adaptive}. Furthermore, we do not need to include models which have been outperformed by our chosen baselines like GRU4REC and RUM \citep{hidasi,chen2018sequential}. For methods that employ similar architectures, we include one of them. ATRank, SASRec and BST \citep{kang,atrank,chen2019behavior} use self-attention to model sequences and we only compare against SASRec. The chosen models are as follows:
\begin{itemize}[leftmargin=*]
\setlength{\itemsep}{0pt}
	\item \textbf{YouTube DNN}. YouTube DNN uses average pooling to integrate behavior embeddings to fixed-width vectors as the user's interest representation \citep{Covington}. 
	\item \textbf{DIN}. DIN proposes the target attention mechanism to soft-search user sequential behaviors with respect to the target item \citep{din}. 
	\item \textbf{DIEN}. DIEN integrates GRU with the target attention mechanism to model user interest evolutions \citep{dien}.
	\item \textbf{SASRec}. SASRec is a self-attentive model based on Transformer \citep{kang}. 
	\item \textbf{MIMN}. MIMN uses a fixed number of memory slots to represent user interests. When a new click takes place, it updates the user memory slots with the GRU-based controller \citep{MIMN}. 
	\item \textbf{UBR4CTR}. UBR4CTR is a two-stage method. The first stage retrieves relevant user behaviors from the sequence with a learnable search method. The second stage feeds retrieved behaviors into a DIN-based deep model \citep{ubr4ctr}. The Amazon datasets contain no item side information, therefore we use a strengthened version of sequence selection for the first stage with multi-head attention on the sequence itself. 
	\item \textbf{SAM 2P/3P}. SAM models without the Memory Enhancement module. 2P refers to 2 iterations of the memory update process, and 3P refers to 3 iterations. To ensure fair comparisons against other methods, we have removed the positional and timestamp encodings for the Books and the Movies datasets.
	\item \textbf{SAM 3P+}. 3 memory update iterations, with the Memory Enhancement module. The number of steps $t$ is 3 for the Memory Enhancement module. 
	\item \textbf{SAM 3P+ts}. SAM 3P+ with timestamp and positional encodings.
\end{itemize}
To ensure the comparison is fair, we remove both the timestamp and positional encodings in SAM 3P and SAM 3P+. This is because we do not include the timestamp and positional encodings in the compared models. The experimental result discussions also revolve around SAM 3P against the compared models. SAM 3P+ts is the full architecture with timestamp and positional encodings.

\subsection{Experimental Results}
We report model performances on three datasets with maximum affordable sequence lengths in Table \ref{tab:auc_max}. Furthermore, we summarize model performances with varying sequence lengths 50, 100, 200, 500 and 1000 in Table \ref{tab:auc_equal}. We have the following important findings:
\begin{itemize}[leftmargin=*]
\setlength{\itemsep}{0pt}
\setlength{\parskip}{0pt}
\item \textit{SAM 3P consistently outperforms compared methods over three datasets.} This demonstrates the effectiveness of our proposed methodology, modeling intra-sequence dependencies and target-sequence dependencies simultaneously. SAM 3P+, with the Memory Enhancement module, has additional improvements over SAM 3P. SAM 3P+ts has limited improvement over SAM 3P+, testifying that the behavior sequence is not strictly ordered in recommender systems. 
\item \textit{SAM 3P constantly outperforms SASRec, the Transformer-based sequential recommender, over equal sequence lengths.} As seen in Table \ref{tab:auc_equal}, SAM 3P outperforms the compared models over equal sequence lengths. Noticeably, SAM 3P outperforms SASRec significantly. This does make sense, considering that SASRec encodes the sequence with multi-head attention with no knowledge on the target item. In contrast, SAM is aware of the target item throughout the encoding process. The ablation study in Section 5.4 verifies the benefits to cross the sequence items and the target item at the very bottom layer of the network. This shows that for recommender systems, modeling the relations between the sequence and the target item is crucial. Even though SASRec explicitly models dependencies between each pair of items in the sequence, its performance is not comparable to methods that model dependencies with the target item.
\item \textit{In general, methods that emphatically perform sequential updates seem to have moderate performance gain}. DIEN uses attention-based GRUs to update the user interest with each sequence item. Similarly, MIMN sequentially updates the nearest user memory slots with each sequence item. Table \ref{tab:auc_equal} shows that DIEN and MIMN constantly outperforms DIN, though the improvement could be moderate in certain experiments. This validates that in recommender systems, the sequential order is not strict. 
\end{itemize}

\begin{table}[ht]
\small
\resizebox{.497\textwidth}{!}{
\begin{tabular}{c|c|c|c}
\toprule
 \hline
 \multicolumn{1}{c|}{}  &\multicolumn{3}{c}{\textbf{AUC (mean$\pm$std)}} \\
 \hline
\multicolumn{1}{c|}{}  &\multicolumn{1}{c|}{ \textbf{Books}} &\multicolumn{1}{c|}{\textbf{Movies}} &\multicolumn{1}{c}{\textbf{Industrial}}
\\  \hline
YouTube &$0.83738(\pm0.00131)$  &$0.83432(\pm0.00164)$ &$0.73534(\pm0.000081)$  \\
DIN             &$0.85162(\pm0.00272)$ &$0.86026(\pm0.00130)$&$0.73749 (\pm0.000126)$ \\
DIEN             &$0.85498(\pm0.00128)$  &$0.86542(\pm0.00072)$ &$0.73807(\pm0.000093)$  \\
SASRec             &$0.82144(\pm0.00748)$ &$0.83690(\pm0.00953)$ &$0.73461(\pm0.000140)$  \\
MIMN             &$0.85228(\pm0.00138)$ &$0.87140(\pm0.00085)$&$0.73678(\pm0.000201)$  \\
UBR4CTR             &$0.84834(\pm0.00062)$ &$0.85957(\pm0.00145)$ &$0.73649(\pm0.000096)$\\
SAM 2P             &$0.85370(\pm0.00196)$ &$0.88214(\pm0.00138)$&$0.73939(\pm0.000034)$ \\
\textbf{SAM 3P}             &$\bm{0.86723(\pm0.00077)}$  &$\bm{0.88352(\pm0.00149)}$ &$\bm{0.74152(\pm0.000093)}$  \\
\hline
SAM 3P+             &$0.86926(\pm0.00142)$  &$0.88628(\pm0.00097)$ &$0.74234(\pm0.000087)$  \\
SAM 3P+ts             &\bf{$0.86997(\pm0.00113)$} &\bf{$0.88714(\pm0.00157)$} &\bf{$0.74238(\pm0.000103)$}  \\
\hline
\bottomrule
\end{tabular}}
\caption{Model performance (AUC) for two public benchmarks and the industrial dataset with maximum affordable sequence lengths. }\label{tab:auc_max}
\end{table}

\begin{table*}[ht]
\small
\resizebox{.99\textwidth}{!}{
\begin{tabular}{ll|lllllll}
\toprule \hline
 &  & \multicolumn{1}{c}{YouTube} & \multicolumn{1}{c}{DIN} &\multicolumn{1}{c}{DIEN} & \multicolumn{1}{c}{SASRec}& \multicolumn{1}{c}{MIMN} &
 \multicolumn{1}{c}{UBR4CTR} & \multicolumn{1}{c}{\textbf{SAM 3P}} \\\hline
 \hline
\multirow{5}{*}{\textbf{Books Dataset}}  & SeqLen=50 & 0.80841 & 0.81873 & 0.84541 & 0.81008 & 0.82753 & 0.81762 & \textbf{0.85662}  \\
 & SeqLen=100 & 0.81729 & 0.84569 & 0.84866 & 0.82144 & 0.84393 & 0.82833 & \textbf{0.86056}  \\
 & SeqLen=200 & 0.82544 & 0.84724 & 0.85498 & N.A. & 0.85228 & 0.83488 & \textbf{0.86377}  \\
  & SeqLen=500 & 0.83252 & 0.84807 & N.A. & N.A. & N.A. & 0.84165 & \textbf{0.86538}  \\
  & SeqLen=1000 & 0.83738& 0.85162 & N.A. & N.A. & N.A. & 0.84834 & \textbf{0.86723}  \\
  \hline
  \multirow{5}{*}{\textbf{Movies Dataset}}  & SeqLen=50 & 0.81336 & 0.83538 & 0.84946 & 0.82978 & 0.85312 & 0.82824 & \textbf{0.86347}  \\
 & SeqLen=100 & 0.82293 & 0.84676 & 0.85997 & 0.83690 & 0.86638 & 0.84297 & \textbf{0.87032}  \\
 & SeqLen=200 & 0.82743 & 0.84917 & 0.86542 & N.A. & 0.87140 & 0.84739 & \textbf{0.87691}  \\
  & SeqLen=500 & 0.83075 & 0.85563 & N.A. & N.A. & N.A. & 0.85301 & \textbf{0.87950}  \\
  & SeqLen=1000 & 0.83432 & 0.86026 & N.A. & N.A. & N.A. & 0.85957 & \textbf{0.88352}  \\
  \hline
  \multirow{5}{*}{\textbf{Industrial Dataset}}  & SeqLen=50 & 0.73019 & 0.73298 & 0.73304 & 0.73296 & 0.73212 & 0.73287 & \textbf{0.73443}  \\
 & SeqLen=100 & 0.73236 & 0.73327 & 0.73331 & 0.73325 & 0.73379 & 0.73309 & \textbf{0.73511}  \\
 & SeqLen=200 & 0.73264 & 0.73331 & 0.73599 & 0.73461 & 0.73678 & 0.73315 & \textbf{0.73796}  \\
  & SeqLen=500 & 0.73371 & 0.73728 & 0.73807 & N.A. & N.A. & 0.73586 & \textbf{0.74029}  \\
  & SeqLen=1000 & 0.73534 & 0.73749 & N.A. & N.A. & N.A. & 0.73649 & \textbf{0.74152}  \\

\hline
\bottomrule
\end{tabular}}
\caption{Model performance (AUC) for varying sequence lengths for the proposed solution and the compared models. Experiments with N.A. incur Out-of-Memory (OOM) error during training.}\label{tab:auc_equal}
\end{table*}

\subsection{Ablation Study}
We conduct ablation study about the model structure and report the results in  Table \ref{tab:abtable}. We remove the timestamp and positional encodings and the Memory Enhancement module to produce the ablation model SAM(w/o. m.e.). We further remove the cross with the target item to produce the ablation model SAM(delayed cross). We replace the element-wise subtraction operation with another element-wise multiplication operation for the ablation model SAM(w/o. subtraction op.). We remove the iterative update process to produce SAM(w/o. iterative walk), which is essentially a DIN-based model. In SAM, the attention mechanism uses the feed-forward attention operator. We replace the feed-forward attention operator in SAM(w/o. iterative walk) to scaled dot-product attention to produce SAM (dot product) where the feature vector in the point-wise dual-query attention is $\boldsymbol{\alpha}_j (\boldsymbol{e}_{j}^{u}, \boldsymbol{v}^T, \boldsymbol{m}_t) = [\boldsymbol{e}_{j}^{u} \odot \boldsymbol{m}_t \,||\, \boldsymbol{e}_{j}^{u} \odot \boldsymbol{v}^T ]$ where $\odot$ represents the dot-product operator. We replace attention with average pooling to produce SAM (w/o. attention), which is to YouTube DNN. The following are our findings: 
\begin{itemize}[leftmargin=*]
\setlength{\itemsep}{0pt}
\setlength{\parskip}{0pt}
\item \textit{The iterative update process models intra-sequence dependencies, which benefits the model performance significantly.} SAM(w/o. m.e.) has a large improvement over SAM(w/o. iterative walk), on par with the improvement of SAM(w/o. iterative walk) over SAM(w/o. attention). It shows it is effective to model intra-sequence dependencies in addition to target-sequence dependencies. 
\item \textit{Delayed cross with the target item results in performance degradation}. SAM(w/o. m.e.) outperforms SAM(delayed cross) by a large extent, showing that crossing the user sequence and the target item at the bottom layer of the network results in performance gain. This also explains for SAM's performance improvements over SASRec, which crosses the Transformer-encoded user sequence and the target item at the very top.
\item \textit{Using feed-forward attention operators results in higher performance than scaled dot-product attention for long sequences.} SAM (w/o. iterative walk) outperforms SAM (dot product). This implies for long sequences, using feed-forward attention operators results in performance gain over scaled dot-product attention. 
\item \textit{Multiple distance measures benefit model performances.} SAM(full) outperforms SAM(w/o. subtraction op.) to a certain extent, showing that multi-faceted distance modeling is beneficial. 
\end{itemize}

\begin{table}[ht]
\small
\resizebox{.477\textwidth}{!}{
\begin{tabular}{cccc}
\toprule
\hline
\multicolumn{1}{c}{\bf Method}  &\multicolumn{3}{c}{\bf AUC(mean$\pm$std)} 
\\
\cline{2-4} 
 & \textbf{Books} &  \textbf{Movies}  & \textbf{Industrial}   \\ 
 \hline
w/o. attention &$0.83738(\pm0.00131)$ &$0.83432(\pm0.00164)$ &$0.73534(\pm0.000081)$  \\
w/o. iterative walk &$0.85162(\pm0.00272)$ &$0.86026(\pm0.00130)$ &$0.73749 (\pm0.000126)$ \\
dot product &$0.84885(\pm0.00147)$ &$0.85644(\pm0.00115)$ &$0.73677(\pm0.000085)$ \\
w/o. subtraction op. &$0.86491(\pm0.00068)$ &$0.87536(\pm0.00133)$ &$0.74020(\pm0.000124)$ \\
delayed cross  &$0.85343(\pm0.00121)$&$0.86037(\pm0.00107)$ &$0.73892(\pm0.000132)$\\
w/o. m.e.  &$0.86723(\pm0.00077)$ &$0.88352(\pm0.00149)$ &$0.74152(\pm0.000093)$  \\
\hline
full (SAM 3P+ts)             &$0.86997(\pm0.00113)$  &$0.88714(\pm0.00157)$ &$0.74238(\pm0.000103)$  \\
\hline
\bottomrule
\end{tabular}
}
\caption{Ablation study on the SAM model structure}\label{tab:abtable}
\end{table}

\subsection{Model Analysis}
We analyze the compared models and summarize the complexity, minimum number of sequential operations, maximum path lengths and encoding paradigms in Table \ref{tab:model-analysis}, with the observations below: 
\begin{itemize}[leftmargin=*]
\setlength{\itemsep}{0pt}
\setlength{\parskip}{0pt}
\item \textit{SAM is efficient with $O(L\cdot d)$ complexity and $O(1)$ number of sequential operations.} SAM incurs $O(L\cdot d)$ complexity. As shown in Section 5.9, the optimal number of memory update iterations is 3, which is a constant, hence the complexity only scales linearly with the sequence length. SASRec is based on Transformer and incurs $O(L^2\cdot d)$ complexity. The minimum number of sequential operations measures the amount of parallelizable computations. Pure attention-based methods are at $O(1)$, while recurrence-based methods are at $O(L)$. Both DIEN and MIMN use sequential update operations per incoming item thus the number of sequential operations is $O(L)$. The number of sequential updates for SAM is $O(1)$ because it does not employ recurrence for each sequence item. GRU is only used for the memory update mechanism, which only needs 3 iterations. Maximum path length refers to the maximum length of signal traversal paths. Research has shown that the length of paths signals need to traverse is the key influence on the ability to learn long-range dependencies \citep{transformer}. The shorter the paths between any combination of positions, the easier it is to learn long-range dependencies. The self-attention mechanism reduces the maximum path length into $O(1)$ since it considers pairwise dependencies between elements in the sequence. The maximum path length for SAM is also $O(1)$ since the dual-query attention queries the sequence items with the memory vector, which contains the sequence information. There is no intra-sequence signal passing in DIN and UBR4CTR. 
\item \textit{SAM has a $\left( CROSS,ENC \right)$ structure, which benefits model performances.} We denote the cross between the user sequence and the target item as $CROSS$ and sequence encoding as $ENC$. The encoding paradigms for DIN and UBR4CTR are $\left( CROSS\right)$, with no sequence encoding after the target attention cross. DIEN is $\left( ENC,CROSS \right)$, encoding the sequence with GRU before target attention. SASRec encodes the sequence with Transformer first and is $\left( ENC,CROSS \right)$. MIMN follows the $\left( ENC,CROSS \right)$ paradigm, using a GRU-based controller to update memory slots with the sequence and crossing the target item at the top. In contrast, SAM has a $\left( CROSS,ENC \right)$ structure. The user sequence crosses with the target item before sequence encoding. Ablation study in Table \ref{tab:abtable} validates empirically that the crossing the user sequence and the target item at the bottom benefits model performances. 
\end{itemize}

\begin{table}[ht]
\small
\resizebox{.477\textwidth}{!}{
\begin{tabular}{c|c|c|c|c}
\toprule
\hline
{\textbf{Method}} &{\textbf{Complexity}} &{\textbf{Seq. Op.}} &{\textbf{Max Path}} &{\textbf{Encoding}} 
\\ \hline
DIN   &$O(L\cdot d)$ & $O(1)$ &$O(\infty)$  &$\left( CROSS \right)$   \\
DIEN&$O(L\cdot d^2)$ &$O(L)$ &$O(L)$ &$\left( ENC,CROSS \right) $\\
SASRec &$O(L^2\cdot d)$ &$O(1)$ &$O(1)$ &$\left( ENC,CROSS \right)$\\
MIMN  &$O(L\cdot d^2)$ &$O(L)$ & $O(L)$ &$\left( ENC,CROSS \right)$ \\
UBR4CTR          &$O(L\cdot d)$ &$O(1)$ &$O(\infty)$ &$\left( CROSS \right)$  \\
SAM      &$O(L\cdot d)$ &$O(1)$ &$O(1)$  &$\left( CROSS,ENC \right)$\\
\hline
\bottomrule
\end{tabular}
}
\caption{Complexity, minimum number of sequential operations(abbreviated as Seq. Op.), maximum path length, and encoding paradigms for compared methods. $L$ is the sequence length and $d$ is the model dimension.}\label{tab:model-analysis}
\end{table}

\subsection{Computational Cost Analysis}
\begin{figure*}
 \begin{subfigure}[t]{.69\columnwidth}
 \hspace{-0.15in}
  \includegraphics[width=\textwidth]{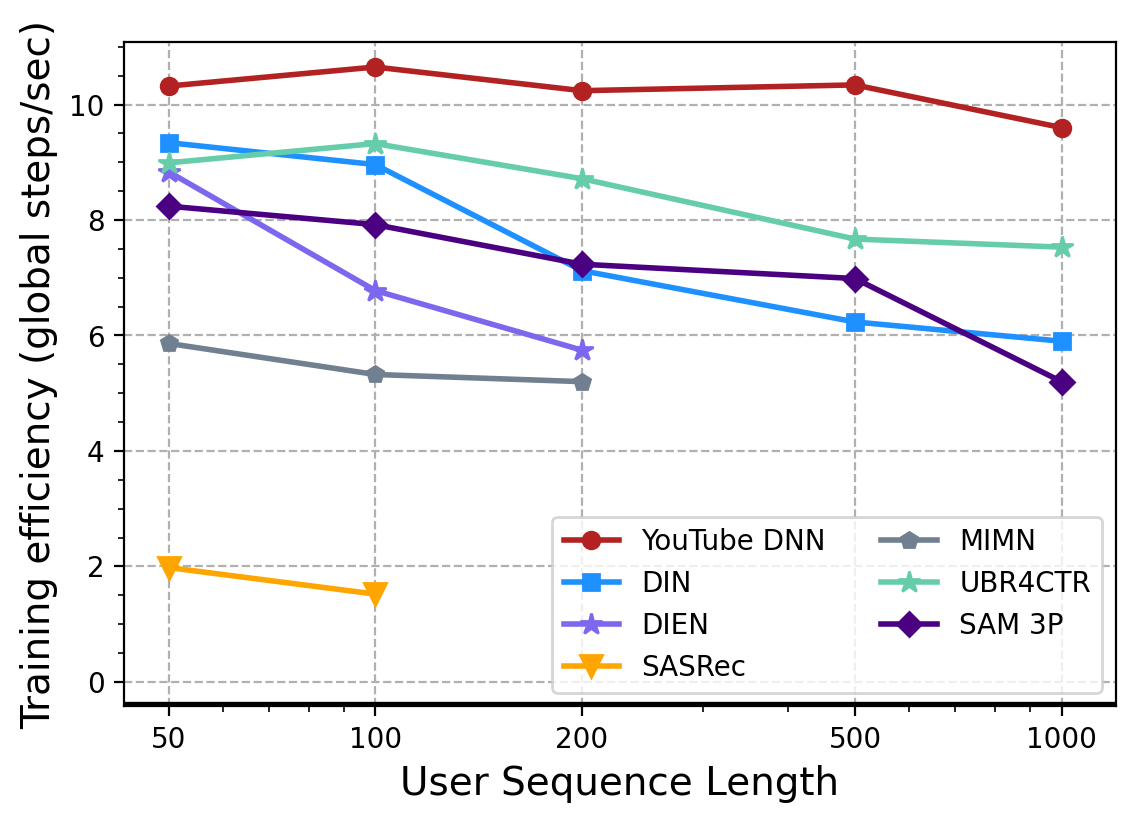}
  \caption{Train Efficiency}
  \label{trainSAM}
  \end{subfigure}
  \begin{subfigure}[t]{.69\columnwidth}
  \hspace{-0.05in}
  \includegraphics[width=\textwidth]{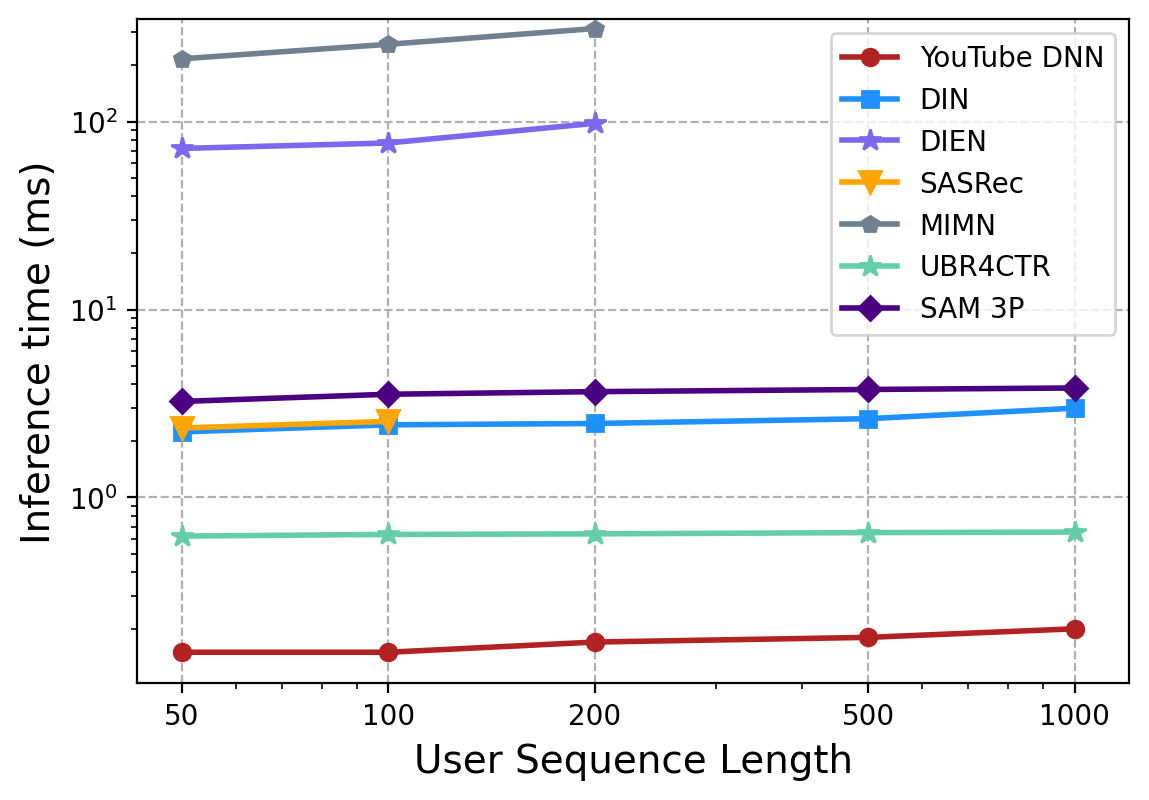}
  \caption{Inference Speed}
  \label{fig:infSAM}
  \end{subfigure}
   \begin{subfigure}[t]{.70\columnwidth}
   \hspace{0.05in}
  \includegraphics[width=\textwidth]{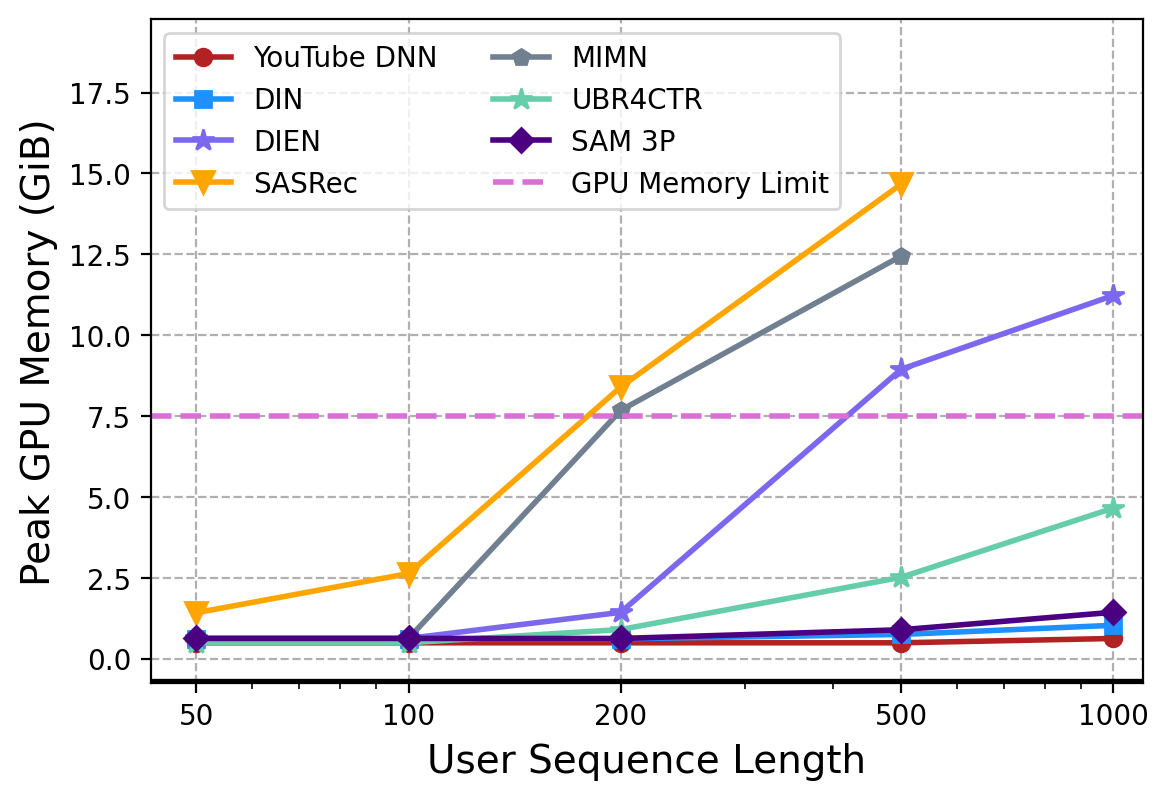}
    \caption{Memory Efficiency}
    \label{fig:memSAM}
  \end{subfigure}
  \caption{Computational cost and memory efficiency for all compared models. The x-axes are on logarithmic scales for all three plots. The y-axis for Fig.\ref{fig:infSAM} is on a logarithmic scale.}
\end{figure*}

We report the run-time for compared methods in Fig.\ref{trainSAM} and the inference time in Fig.\ref{fig:infSAM}. The run-time efficiency is measured by global steps per second during training. The real-time efficiency is measured by the inference time in milliseconds. The x-axis for Fig.\ref{trainSAM} is on a logarithmic scale. Both axes for Fig.\ref{fig:infSAM} are on logarithmic scales. For the inference time, we only measure the forward pass cost, excluding the input encoding cost. We use inputs with lengths 50, 100, 200, 500 and 1000. Experiments with missing data incur Out-of-Memory (OOM) errors that stop training. All experiments are conducted on Tesla A100 GPU with 10GiB memory. We summarize our findings below:
\begin{itemize}[leftmargin=*]
\setlength{\itemsep}{0pt}
\setlength{\parskip}{0pt}
\item \textit{SAM is computationally efficient with increasing sequence lengths.} SAM involves matrix operations heavily, which are highly optimized and parallelizable on GPU. SAM has similar training efficiency as DIN when sequences are at length 1000. The forward pass inference cost at sequence length 1000 is only 3.6ms.
\item \textit{Methods based on sequential updates are computationally expensive for training and inference.} DIEN and MIMN, the two models based on sequential updates, have significantly lower training and inference speeds. The inference time for MIMN at sequence length 50 is 216ms, while the industrial norm is 30ms to 80ms. More analysis shows that MIMN's addressing head is time-consuming. The computational inefficiency forces MIMN to separate the user and the item sides, performing user side inference prior to online scoring. Similarly, DIEN's inference time has reached 100ms at sequence length 200. Unlike MIMN, DIEN cannot separate the user and item sides since it performs target attention on top of the GRU encoder. This limits DIEN's scalability to longer sequences.
\item \textit{SASRec, the self-attentive method, is efficient during inference time, but not during training time.} The inference latency for SASRec is significantly lower compared to methods with sequential updates like DIEN and MIMN. This does make sense since self-attention allows for more parallelizations compared to DIEN and MIMN. Training is relatively slow for SASRec.
\end{itemize}

\subsection{Memory Consumption}
We evaluate the memory efficiency by measuring the peak memory usage in GiB in Fig.\ref{fig:memSAM}. The memory limit is 10GiB. Experiments above the horizontal dotted line in Fig.\ref{fig:memSAM} incur Out-of-Memory (OOM) errors. We summarize our findings below:

\begin{itemize}[leftmargin=*]
\setlength{\itemsep}{0pt}
\setlength{\parskip}{0pt}

\item \textit{SAM is efficient with memory consumption increasing linearly with sequence lengths.} SAM incurs linear space complexity. The memory overhead is the user memory vector, the same size as the target item. The low peak memory consumption at varying lengths testifies its memory efficiency. 
\item \textit{Self-attentive methods have the most memory usage increase with increasing sequence lengths. The NTM-based MIMN is also memory-hungry.} SASRec incurs Out-Of-Memory (OOM) errors on sequences longer than 100. Fig.\ref{fig:memSAM} also shows that its memory consumption increase is the most substantial with increasing sequence lengths, testifying the $O(L^2)$ memory bottleneck. MIMN is also memory-hungry, since keeping additional user memory slots results in memory overheads.
\end{itemize}

\subsection{Extremely Long Sequences}
To analyze computational and memory efficiencies for even longer sequences, we use synthetic inputs with varying lengths from 1K to 16K. The experimental settings are the same as in Section 5.1. We report the forward pass inference time in Fig.\ref{fig:longInf} and the memory statistics in Fig.\ref{fig:longMem}. We experiment on YouTube DNN, DIN and SAM since UBR4CTR uses a DIN-based model for the second stage and the other compared methods cannot afford sequences beyond length 1000. We summarize our findings below:

\begin{figure}
  \centering
  \begin{subfigure}{.495\columnwidth}
  \includegraphics[width=\textwidth]{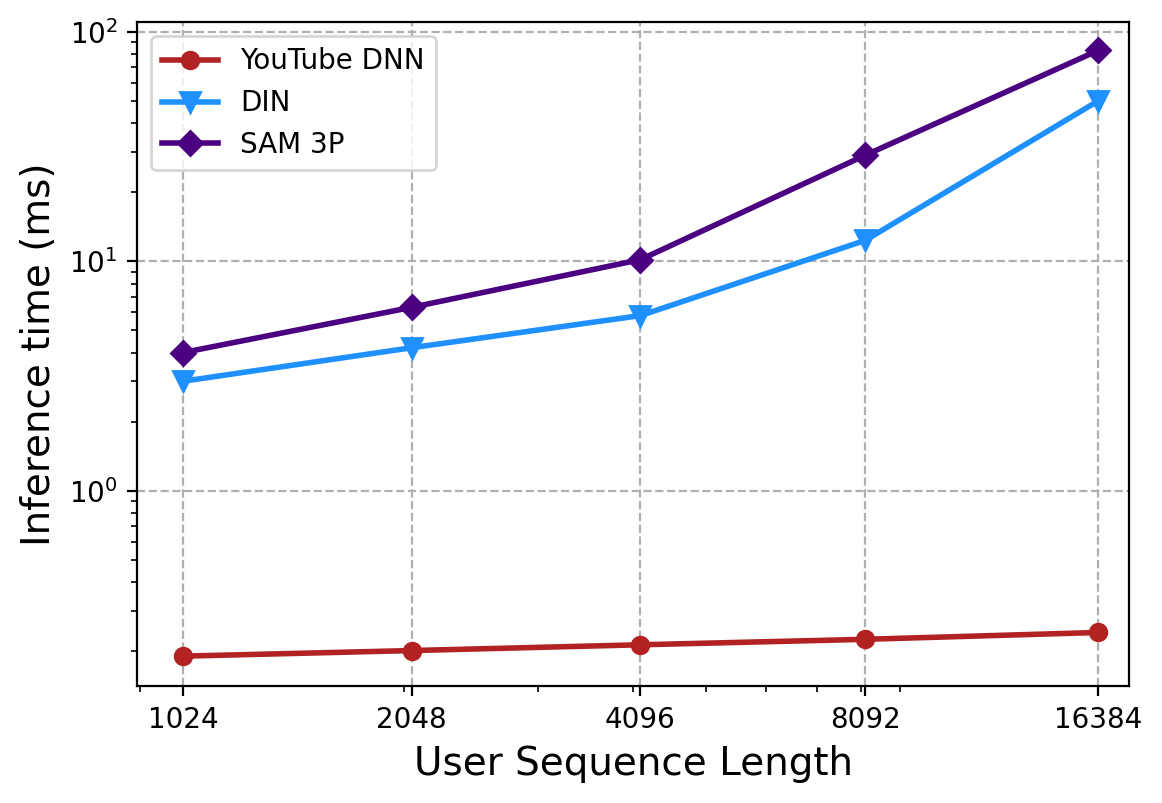}
  \caption{Inference Speed}
  \label{fig:longInf}
  \end{subfigure}
  \begin{subfigure}{.495\columnwidth}
  \includegraphics[width=\textwidth]{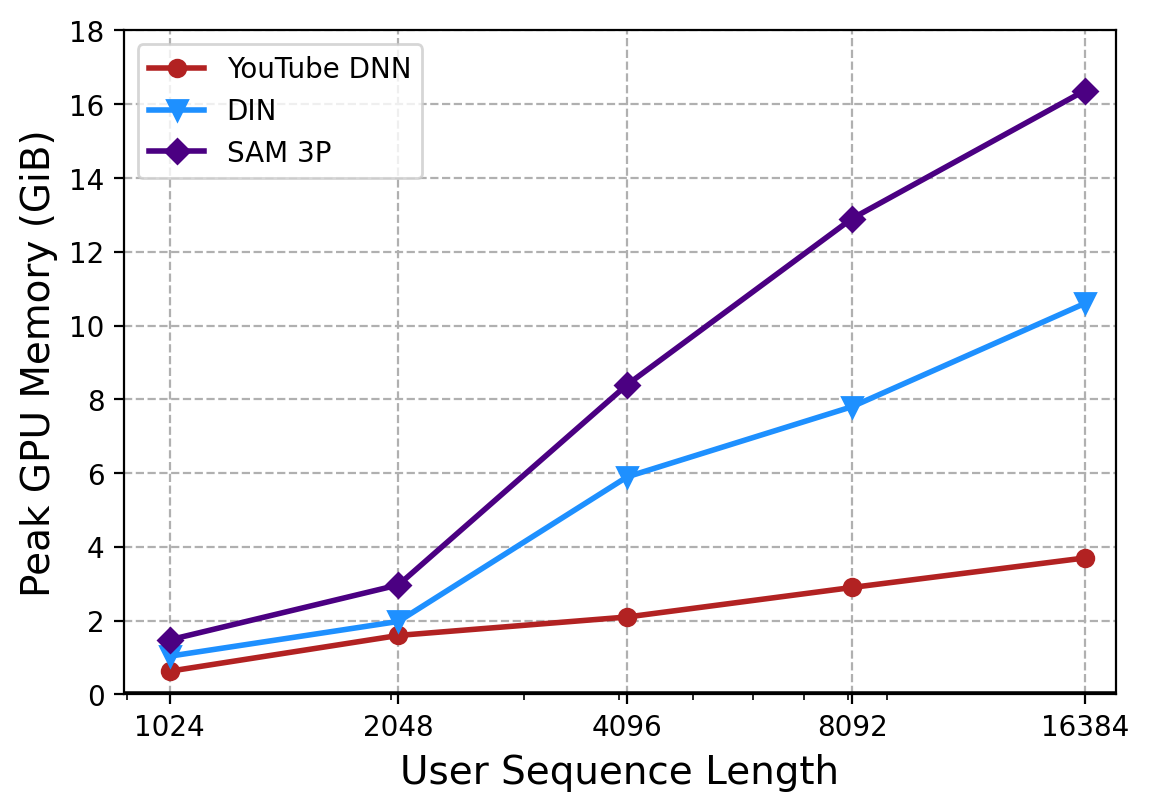}
  \caption{Memory Efficiency}
  \label{fig:longMem}
  \end{subfigure}
  \caption{Inference time and peak memory usage for extremely long sequences with lengths up to 16K. The y-axis for the inference time is on a logarithmic scale.}
\end{figure}

\begin{itemize}[leftmargin=*]
\setlength{\itemsep}{0pt}
\setlength{\parskip}{0pt}
\item \textit{The computational costs for SAM are affordable for very long sequences under GPU environments.} The forward pass cost is within 80ms for SAM on sequences of length 16K. SAM relies heavily on matrix operations, which are highly optimized to be parallelizable on GPU. The inference time is only 1.7x in comparison to DIN. The increase in inference latency is a trade-off with the added ability to model intra-sequence dependencies. Since the ablation study in Section 5.4 shows the performance improvement with modeling intra-sequence dependencies is large, the inference latency increase is relatively insignificant. 

\item \textit{Memory costs are not limiting SAM's scalability to even longer sequences.} As seen in Fig.\ref{fig:longMem}, SAM's peak memory consumption is 16GiB when the sequence length reaches 16K. The memory consumption is only about 1.6x relative to the memory consumption for DIN. It empirically verifies that the linear memory complexities for SAM and DIN allow for their scalability to extremely long sequences. The quadratic memory complexity is indeed a major bottleneck for self-attention based methods.

\end{itemize}

\subsection{Sensitivity w.r.t Number of Memory Update Iterations}
\begin{figure}
  \centering
  \includegraphics[width=.495\columnwidth]{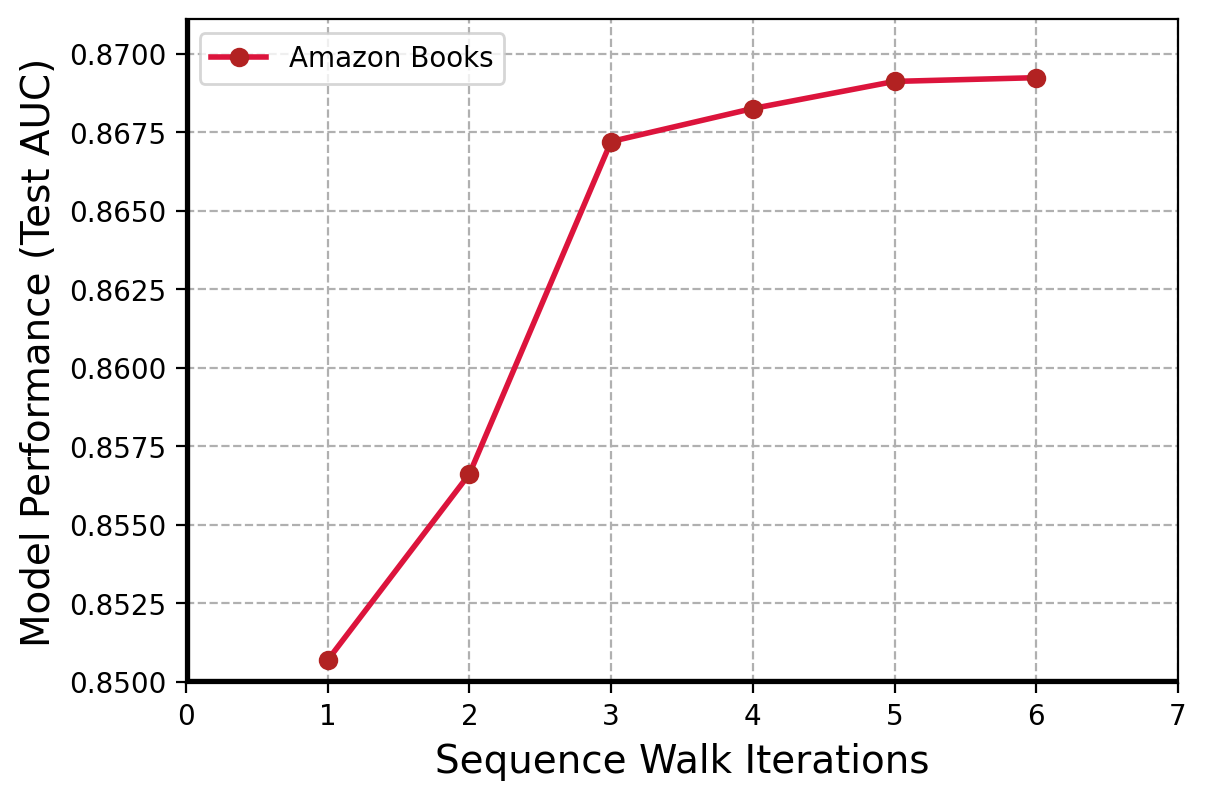}
  \includegraphics[width=.495\columnwidth]{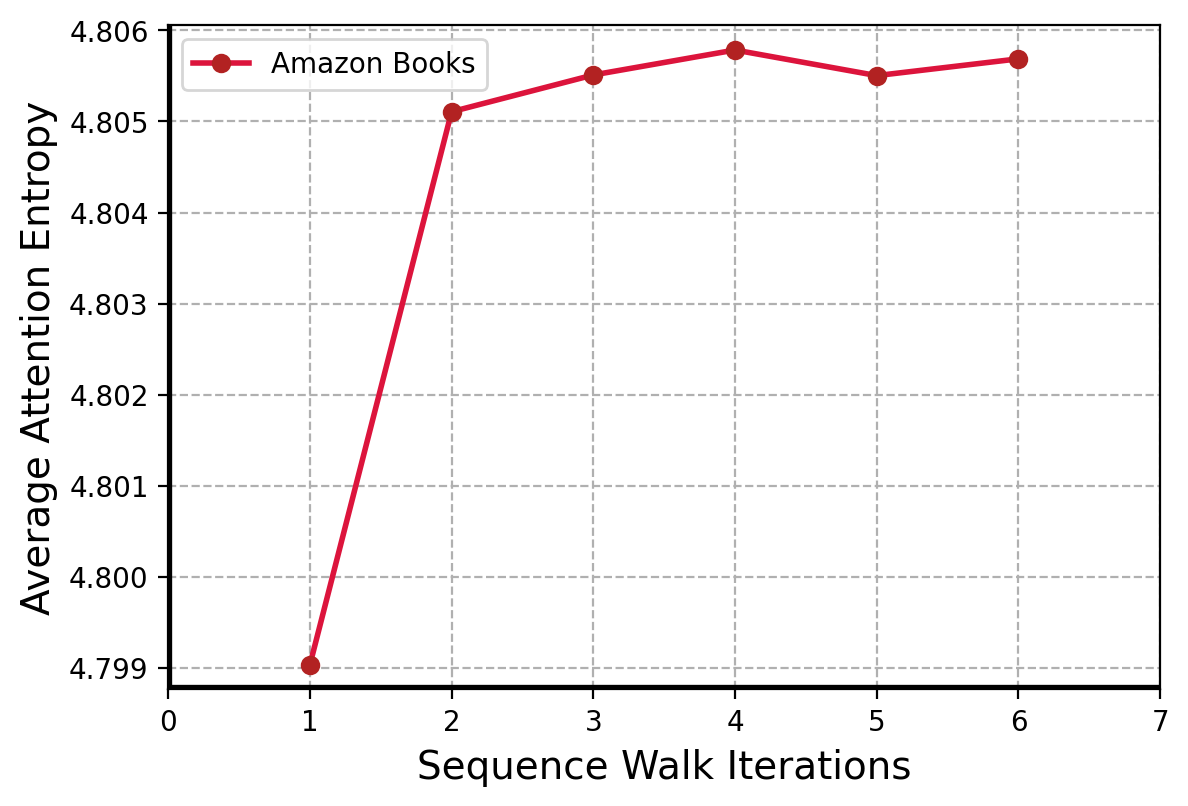}
\includegraphics[width=.495\columnwidth]{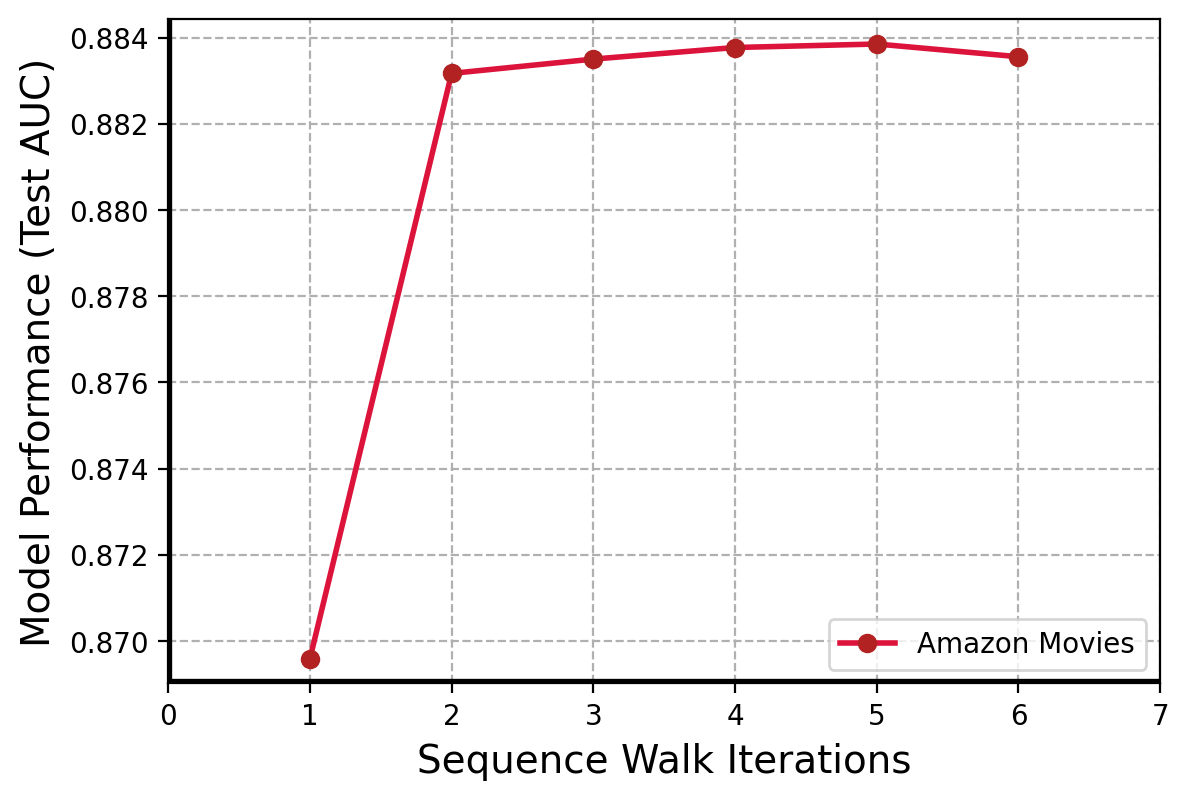}
  \includegraphics[width=.495\columnwidth]{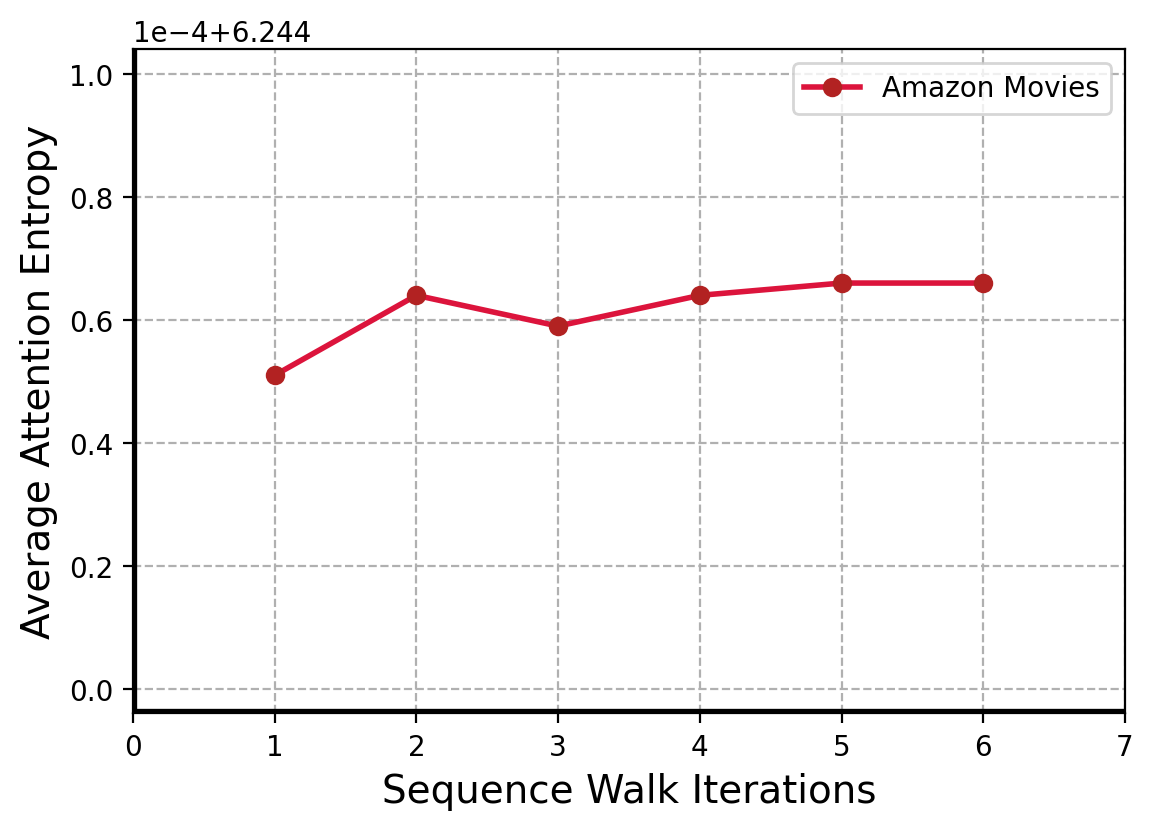}
\includegraphics[width=.495\columnwidth]{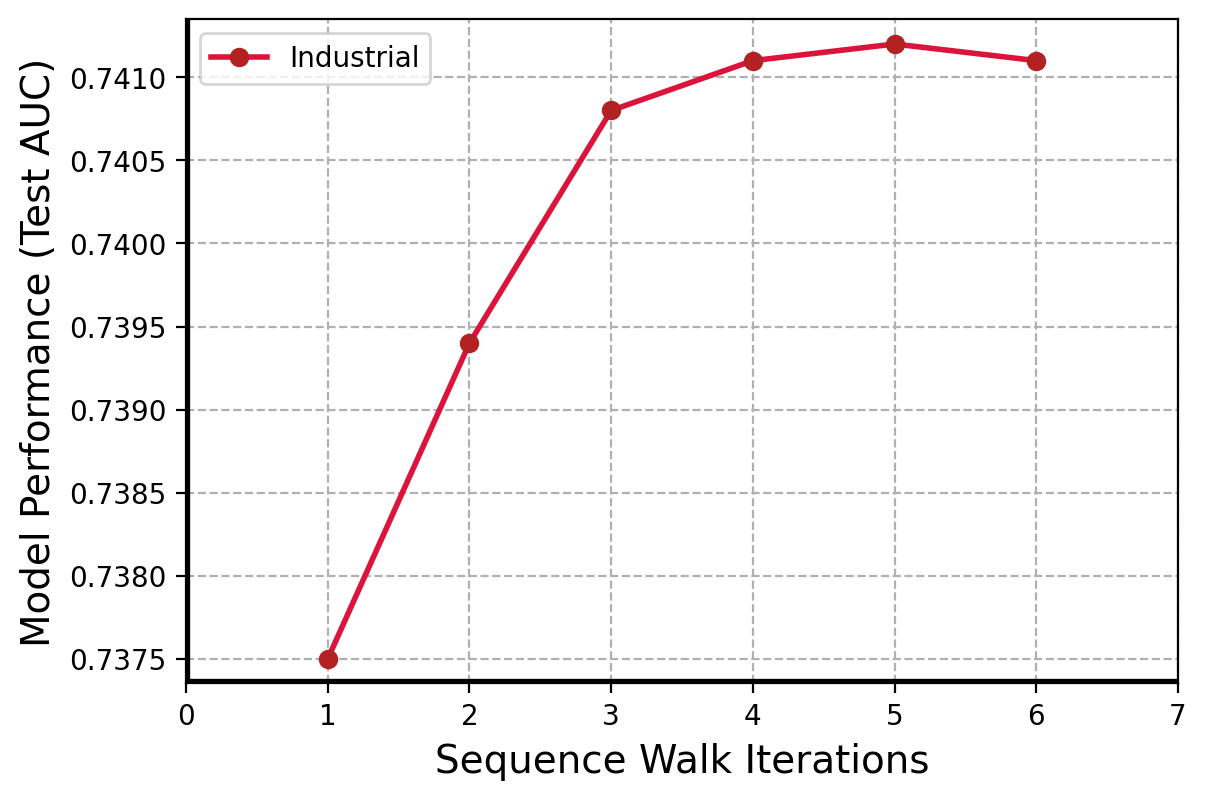}
  \includegraphics[width=.495\columnwidth]{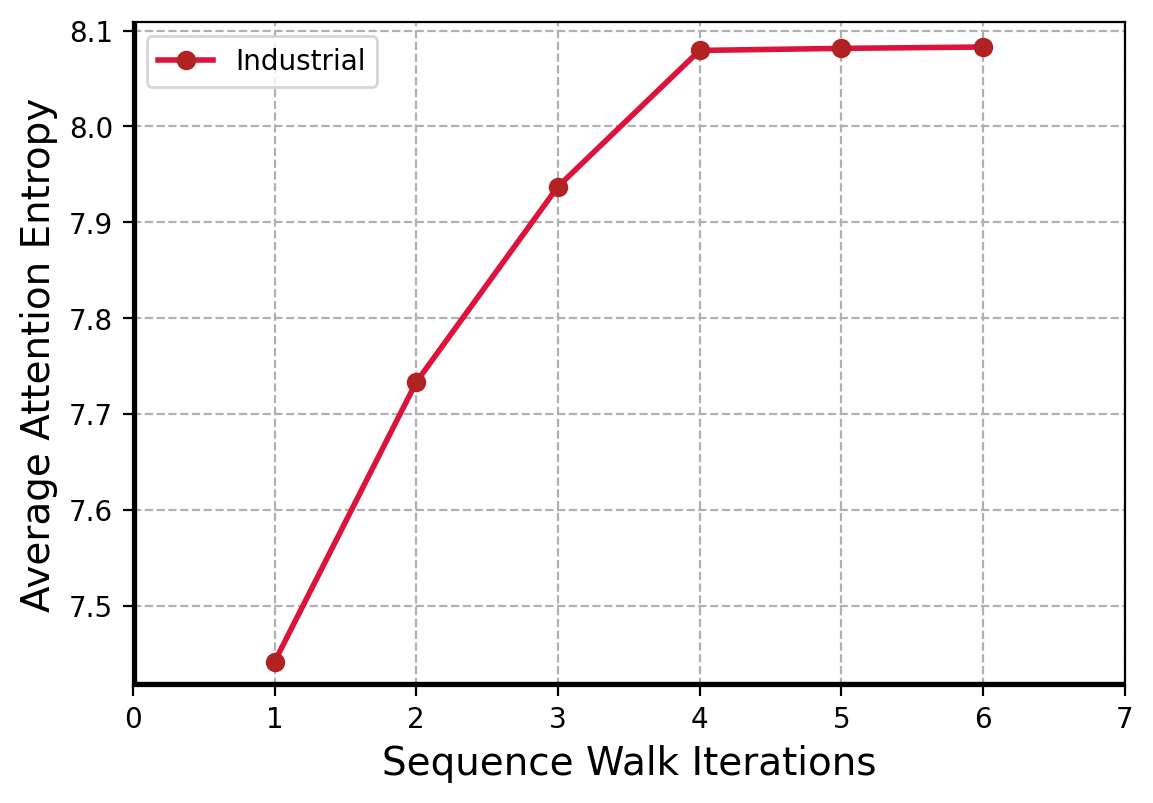}
  \caption{Model performance (AUC) and the entropy of the attention distribution against memory update iterations.}
  \label{auciter}
\end{figure}

We investigate the impact of memory update iterations. The number of memory update iterations equals the number of sequence walks since the memory is updated after a full pass of the sequence. Fig.\ref{auciter} shows SAM's AUC performance against sequence walk iterations. For all three datasets, the AUC increases with more iterations and stabilizes after 3 to 4 iterations, showing that the optimal hyper-parameter for the memory update mechanism is 3. The AUC increase is large for the first two iterations, showing that the first two memory update iterations result in most performance gain. \\ 
In order to visualize the degree of the attention dispersion, we calculate the entropy of the attention distribution \citep{vig2019analyzing,ghader2017does}:
\begin{equation}
 Entropy_{\alpha}(x) =  -\sum_i^L(\alpha_i(x)  \log({\alpha_i(x))})
\label{entropy_eqn}
\end{equation}
where $\alpha_i(x)$ is the normalized attention score for position $i$. The attention entropy is averaged over samples and plotted in Fig.\ref{auciter}. The entropy also stabilizes after 3 to 4 iterations. \\
Both trends show that there are negligible benefits with additional iterative memory updates beyond 3 to 4 iterations, which justifies that the optimal iteration hyper-parameter is 3. \\


\section{Online A/B Performance}
We have deployed the proposed solution on one of the largest international E-commerce platforms for item recommendation. From 2021-10-15 to 2021-11-30, we conduct strict and thorough online A/B test experiments to validate the proposed SAM model. The baseline is the last deployed production model, a DIN-based deep model with sequences truncated to the most recent 50 user behaviors. SAM is implemented on user click sequences with length 1000, keeping other components exactly the same. Table \ref{tab:online} summarizes the A/B test results. Besides the canonical CTR metric, Total Clicked Items Count (TCIC) refers to the total number of distinct items having at least 1 click. Clicked Categories Count (CCC) refers to the average number of categories clicked per user. TCIC and CCC are diversity measures for recommender systems. With more items being clicked and more categories clicked by each user, the recommender system has a higher diversity. As seen in Table \ref{tab:online}, SAM improves CTR by 7.30\%, TCIC by 15.36\% and CCC by 7.19\%. \\ 


\begin{table}[t]
\caption{Online A/B test results for consecutive 9 days. The row Impr denotes relative improvement.}

\begin{tabular}{c|c|c|c}
\toprule
\hline
\multicolumn{1}{c|}{}  &\multicolumn{3}{c}{\textbf {Online A/B Metrics (mean$\pm$std)}} \\  \hline
\multicolumn{1}{c|}{}  &\multicolumn{1}{c|}{\bf CTR } &\multicolumn{1}{c|}{\bf TCIC} 
&\multicolumn{1}{c}{\bf CCC} 
\\ \hline
Base &4.4254\%$\pm$0.0244\% &325741$\pm$2701.9 &2.979$\pm$0.0133 \\
SAM &4.7482\%$\pm$0.0222\% & 375733$\pm$4055.1 &3.193$\pm$0.0181 \\
\textbf{Impr}  &\textbf{7.30\%$\pm$0.93\%} & \textbf{15.36\%$\pm$1.65\%} &\textbf{7.19\%$\pm$0.80\%} \\
\hline
\bottomrule
\end{tabular}
\label{tab:online}
\end{table}

\section{Deployment to Production}
Since December 2021, we have deployed SAM on all the traffic of the main page of one of the largest international E-commerce platforms, hosting 20 million daily users with a traffic volume of 1500 QPS (Query Per Second). To deploy complex models on industrial recommender systems requires great effort. The two most critical challenges we have tackled are latency and storage constraints. 
\begin{itemize}[leftmargin=*]
\setlength{\itemsep}{0pt}
\setlength{\parskip}{0pt}
\item \textbf{Latency Constraints.} The typical upper limit for real-time industrial recommender response time is 30ms to 80ms. When we first deploy SAM on CPU clusters, the real-time inference time exceeds 300ms. SAM relies on matrix computations heavily. Since matrix computations are extensively researched and highly optimized on GPU\citep{gpu1, gpu2, gpu3, gpu4}, we deploy SAM on GPU clusters. We use 48 Nvidia Tesla A100 GPUs to serve the traffic volume of 1500 QPS. The inference time is within 30ms.
\item \textbf{Storage Constraints.} The storage constraints refer to both the storage space to store the offline samples and that to store the user sequences for online inference. With a 0.1 sample rate on the negative samples, the sample size for 1-day sample is 60 million and the storage volume is 1 terabyte (TB). We keep samples for 45 days, which account for a total storage size of 45 terabytes. When the model is served online, we need to feed the user sequences. We use the internal online graph storage system, with a total storage of 350 gigabytes(GB).

\end{itemize}

\section{Conclusion}
In this paper, we propose a novel user sequential behavior model, SAM, which models long sequences with lengths on the scale of thousands. It can model intra-sequence dependencies and target-sequence dependencies within $O(L)$ complexity and $O(1)$ number of sequential operations. Empirical results on several datasets demonstrate its effectiveness in modeling both long user behavior sequences and short sequences. SAM supports efficient training and real-time inference. It is deployed successfully on an E-commerce recommender system with 1500 QPS, with a significant improvement of 7.30\% CTR over the DIN-based industrial baseline. \\

\section{Citations and Bibliographies}

\bibliographystyle{ACM-Reference-Format}
\bibliography{sample-base}


\begin{thebibliography}{51}


\ifx \showCODEN    \undefined \def \showCODEN     #1{\unskip}     \fi
\ifx \showDOI      \undefined \def \showDOI       #1{#1}\fi
\ifx \showISBNx    \undefined \def \showISBNx     #1{\unskip}     \fi
\ifx \showISBNxiii \undefined \def \showISBNxiii  #1{\unskip}     \fi
\ifx \showISSN     \undefined \def \showISSN      #1{\unskip}     \fi
\ifx \showLCCN     \undefined \def \showLCCN      #1{\unskip}     \fi
\ifx \shownote     \undefined \def \shownote      #1{#1}          \fi
\ifx \showarticletitle \undefined \def \showarticletitle #1{#1}   \fi
\ifx \showURL      \undefined \def \showURL       {\relax}        \fi
\providecommand\bibfield[2]{#2}
\providecommand\bibinfo[2]{#2}
\providecommand\natexlab[1]{#1}
\providecommand\showeprint[2][]{arXiv:#2}

\bibitem[\protect\citeauthoryear{Agarwal, Chen, and Elango}{Agarwal
  et~al\mbox{.}}{2009}]%
        {agarwal}
\bibfield{author}{\bibinfo{person}{Deepak Agarwal}, \bibinfo{person}{Bee-Chung
  Chen}, {and} \bibinfo{person}{Pradheep Elango}.}
  \bibinfo{year}{2009}\natexlab{}.
\newblock \showarticletitle{Spatio-temporal models for estimating click-through
  rate}. In \bibinfo{booktitle}{\emph{Proceedings of the 18th international
  conference on World wide web}}. \bibinfo{pages}{21--30}.
\newblock


\bibitem[\protect\citeauthoryear{Bahdanau, Cho, and Bengio}{Bahdanau
  et~al\mbox{.}}{2014}]%
        {bahdanau}
\bibfield{author}{\bibinfo{person}{Dzmitry Bahdanau},
  \bibinfo{person}{Kyunghyun Cho}, {and} \bibinfo{person}{Yoshua Bengio}.}
  \bibinfo{year}{2014}\natexlab{}.
\newblock \showarticletitle{Neural machine translation by jointly learning to
  align and translate}.
\newblock \bibinfo{journal}{\emph{arXiv preprint arXiv:1409.0473}}
  (\bibinfo{year}{2014}).
\newblock


\bibitem[\protect\citeauthoryear{Cecilia, Garc{\'i}a, and Ujald{\'o}n}{Cecilia
  et~al\mbox{.}}{2009}]%
        {gpu1}
\bibfield{author}{\bibinfo{person}{Jos{\'e}~M. Cecilia},
  \bibinfo{person}{Jos{\'e}~M. Garc{\'i}a}, {and} \bibinfo{person}{Manuel
  Ujald{\'o}n}.} \bibinfo{year}{2009}\natexlab{}.
\newblock \showarticletitle{The GPU on the Matrix-Matrix Multiply: Performance
  Study and Contributions}. In \bibinfo{booktitle}{\emph{PARCO}}.
\newblock


\bibitem[\protect\citeauthoryear{Chan, Jaitly, Le, and Vinyals}{Chan
  et~al\mbox{.}}{2015}]%
        {chan}
\bibfield{author}{\bibinfo{person}{William Chan}, \bibinfo{person}{Navdeep
  Jaitly}, \bibinfo{person}{Quoc~V Le}, {and} \bibinfo{person}{Oriol Vinyals}.}
  \bibinfo{year}{2015}\natexlab{}.
\newblock \showarticletitle{Listen, attend and spell}.
\newblock \bibinfo{journal}{\emph{arXiv preprint arXiv:1508.01211}}
  (\bibinfo{year}{2015}).
\newblock


\bibitem[\protect\citeauthoryear{Chaudhari, Mithal, Polatkan, and
  Ramanath}{Chaudhari et~al\mbox{.}}{2021}]%
        {chaudhari}
\bibfield{author}{\bibinfo{person}{Sneha Chaudhari}, \bibinfo{person}{Varun
  Mithal}, \bibinfo{person}{Gungor Polatkan}, {and} \bibinfo{person}{Rohan
  Ramanath}.} \bibinfo{year}{2021}\natexlab{}.
\newblock \showarticletitle{An attentive survey of attention models}.
\newblock \bibinfo{journal}{\emph{ACM Transactions on Intelligent Systems and
  Technology (TIST)}} \bibinfo{volume}{12}, \bibinfo{number}{5}
  (\bibinfo{year}{2021}), \bibinfo{pages}{1--32}.
\newblock


\bibitem[\protect\citeauthoryear{Chen, Zhao, Li, Huang, and Ou}{Chen
  et~al\mbox{.}}{2019}]%
        {chen2019behavior}
\bibfield{author}{\bibinfo{person}{Qiwei Chen}, \bibinfo{person}{Huan Zhao},
  \bibinfo{person}{Wei Li}, \bibinfo{person}{Pipei Huang}, {and}
  \bibinfo{person}{Wenwu Ou}.} \bibinfo{year}{2019}\natexlab{}.
\newblock \showarticletitle{Behavior sequence transformer for e-commerce
  recommendation in alibaba}. In \bibinfo{booktitle}{\emph{Proceedings of the
  1st International Workshop on Deep Learning Practice for High-Dimensional
  Sparse Data}}. \bibinfo{pages}{1--4}.
\newblock


\bibitem[\protect\citeauthoryear{Chen, Xu, Zhang, Tang, Cao, Qin, and Zha}{Chen
  et~al\mbox{.}}{2018}]%
        {chen2018sequential}
\bibfield{author}{\bibinfo{person}{Xu Chen}, \bibinfo{person}{Hongteng Xu},
  \bibinfo{person}{Yongfeng Zhang}, \bibinfo{person}{Jiaxi Tang},
  \bibinfo{person}{Yixin Cao}, \bibinfo{person}{Zheng Qin}, {and}
  \bibinfo{person}{Hongyuan Zha}.} \bibinfo{year}{2018}\natexlab{}.
\newblock \showarticletitle{Sequential recommendation with user memory
  networks}. In \bibinfo{booktitle}{\emph{Proceedings of the eleventh ACM
  international conference on web search and data mining}}.
  \bibinfo{pages}{108--116}.
\newblock


\bibitem[\protect\citeauthoryear{Child, Gray, Radford, and Sutskever}{Child
  et~al\mbox{.}}{2019}]%
        {child}
\bibfield{author}{\bibinfo{person}{Rewon Child}, \bibinfo{person}{Scott Gray},
  \bibinfo{person}{Alec Radford}, {and} \bibinfo{person}{Ilya Sutskever}.}
  \bibinfo{year}{2019}\natexlab{}.
\newblock \showarticletitle{Generating long sequences with sparse
  transformers}.
\newblock \bibinfo{journal}{\emph{arXiv preprint arXiv:1904.10509}}
  (\bibinfo{year}{2019}).
\newblock


\bibitem[\protect\citeauthoryear{Covington, Adams, and Sargin}{Covington
  et~al\mbox{.}}{2016}]%
        {Covington}
\bibfield{author}{\bibinfo{person}{Paul Covington}, \bibinfo{person}{Jay
  Adams}, {and} \bibinfo{person}{Emre Sargin}.}
  \bibinfo{year}{2016}\natexlab{}.
\newblock \showarticletitle{Deep neural networks for youtube recommendations}.
  In \bibinfo{booktitle}{\emph{Proceedings of the 10th ACM conference on
  recommender systems}}. \bibinfo{pages}{191--198}.
\newblock


\bibitem[\protect\citeauthoryear{Cui, Chen, Wei, Wang, Liu, and Hu}{Cui
  et~al\mbox{.}}{2016}]%
        {cui}
\bibfield{author}{\bibinfo{person}{Yiming Cui}, \bibinfo{person}{Zhipeng Chen},
  \bibinfo{person}{Si Wei}, \bibinfo{person}{Shijin Wang},
  \bibinfo{person}{Ting Liu}, {and} \bibinfo{person}{Guoping Hu}.}
  \bibinfo{year}{2016}\natexlab{}.
\newblock \showarticletitle{Attention-over-attention neural networks for
  reading comprehension}.
\newblock \bibinfo{journal}{\emph{arXiv preprint arXiv:1607.04423}}
  (\bibinfo{year}{2016}).
\newblock


\bibitem[\protect\citeauthoryear{Dalton, Baxter, Merrill, Olson, and
  Garland}{Dalton et~al\mbox{.}}{2015}]%
        {gpu3}
\bibfield{author}{\bibinfo{person}{Steven Dalton}, \bibinfo{person}{Sean
  Baxter}, \bibinfo{person}{Duane Merrill}, \bibinfo{person}{Luke Olson}, {and}
  \bibinfo{person}{Michael Garland}.} \bibinfo{year}{2015}\natexlab{}.
\newblock \showarticletitle{Optimizing sparse matrix operations on gpus using
  merge path}. In \bibinfo{booktitle}{\emph{2015 IEEE International Parallel
  and Distributed Processing Symposium}}. IEEE, \bibinfo{pages}{407--416}.
\newblock


\bibitem[\protect\citeauthoryear{Fatahalian, Sugerman, and Hanrahan}{Fatahalian
  et~al\mbox{.}}{2004}]%
        {gpu4}
\bibfield{author}{\bibinfo{person}{Kayvon Fatahalian}, \bibinfo{person}{Jeremy
  Sugerman}, {and} \bibinfo{person}{Pat Hanrahan}.}
  \bibinfo{year}{2004}\natexlab{}.
\newblock \showarticletitle{Understanding the efficiency of GPU algorithms for
  matrix-matrix multiplication}. In \bibinfo{booktitle}{\emph{Proceedings of
  the ACM Siggraph/Eurographics conference on Graphics hardware}}.
  \bibinfo{pages}{133--137}.
\newblock


\bibitem[\protect\citeauthoryear{Ghader and Monz}{Ghader and Monz}{2017}]%
        {ghader2017does}
\bibfield{author}{\bibinfo{person}{Hamidreza Ghader} {and}
  \bibinfo{person}{Christof Monz}.} \bibinfo{year}{2017}\natexlab{}.
\newblock \showarticletitle{What does attention in neural machine translation
  pay attention to?}
\newblock \bibinfo{journal}{\emph{arXiv preprint arXiv:1710.03348}}
  (\bibinfo{year}{2017}).
\newblock


\bibitem[\protect\citeauthoryear{Graves, Wayne, and Danihelka}{Graves
  et~al\mbox{.}}{2014}]%
        {Graves}
\bibfield{author}{\bibinfo{person}{Alex Graves}, \bibinfo{person}{Greg Wayne},
  {and} \bibinfo{person}{Ivo Danihelka}.} \bibinfo{year}{2014}\natexlab{}.
\newblock \showarticletitle{Neural turing machines}.
\newblock \bibinfo{journal}{\emph{arXiv preprint arXiv:1410.5401}}
  (\bibinfo{year}{2014}).
\newblock


\bibitem[\protect\citeauthoryear{Guo, Shi, and Liu}{Guo et~al\mbox{.}}{2020}]%
        {guo2020intention}
\bibfield{author}{\bibinfo{person}{Xueliang Guo}, \bibinfo{person}{Chongyang
  Shi}, {and} \bibinfo{person}{Chuanming Liu}.}
  \bibinfo{year}{2020}\natexlab{}.
\newblock \showarticletitle{Intention modeling from ordered and unordered
  facets for sequential recommendation}. In
  \bibinfo{booktitle}{\emph{Proceedings of The Web Conference 2020}}.
  \bibinfo{pages}{1127--1137}.
\newblock


\bibitem[\protect\citeauthoryear{Hidasi, Karatzoglou, Baltrunas, and
  Tikk}{Hidasi et~al\mbox{.}}{2015}]%
        {hidasi}
\bibfield{author}{\bibinfo{person}{Bal{\'a}zs Hidasi},
  \bibinfo{person}{Alexandros Karatzoglou}, \bibinfo{person}{Linas Baltrunas},
  {and} \bibinfo{person}{Domonkos Tikk}.} \bibinfo{year}{2015}\natexlab{}.
\newblock \showarticletitle{Session-based recommendations with recurrent neural
  networks}.
\newblock \bibinfo{journal}{\emph{arXiv preprint arXiv:1511.06939}}
  (\bibinfo{year}{2015}).
\newblock


\bibitem[\protect\citeauthoryear{Hochreiter and Schmidhuber}{Hochreiter and
  Schmidhuber}{1997}]%
        {Hochreiter}
\bibfield{author}{\bibinfo{person}{Sepp Hochreiter} {and}
  \bibinfo{person}{J{\"u}rgen Schmidhuber}.} \bibinfo{year}{1997}\natexlab{}.
\newblock \showarticletitle{Long short-term memory}.
\newblock \bibinfo{journal}{\emph{Neural computation}} \bibinfo{volume}{9},
  \bibinfo{number}{8} (\bibinfo{year}{1997}), \bibinfo{pages}{1735--1780}.
\newblock


\bibitem[\protect\citeauthoryear{Juan, Zhuang, Chin, and Lin}{Juan
  et~al\mbox{.}}{2016}]%
        {juan2016field}
\bibfield{author}{\bibinfo{person}{Yuchin Juan}, \bibinfo{person}{Yong Zhuang},
  \bibinfo{person}{Wei-Sheng Chin}, {and} \bibinfo{person}{Chih-Jen Lin}.}
  \bibinfo{year}{2016}\natexlab{}.
\newblock \showarticletitle{Field-aware factorization machines for CTR
  prediction}. In \bibinfo{booktitle}{\emph{Proceedings of the 10th ACM
  conference on recommender systems}}. \bibinfo{pages}{43--50}.
\newblock


\bibitem[\protect\citeauthoryear{Kang and McAuley}{Kang and McAuley}{2018}]%
        {kang}
\bibfield{author}{\bibinfo{person}{Wang-Cheng Kang} {and}
  \bibinfo{person}{Julian McAuley}.} \bibinfo{year}{2018}\natexlab{}.
\newblock \showarticletitle{Self-attentive sequential recommendation}. In
  \bibinfo{booktitle}{\emph{2018 IEEE International Conference on Data Mining
  (ICDM)}}. IEEE, \bibinfo{pages}{197--206}.
\newblock


\bibitem[\protect\citeauthoryear{Kingma and Ba}{Kingma and Ba}{2014}]%
        {KingmaB14}
\bibfield{author}{\bibinfo{person}{Diederik~P Kingma} {and}
  \bibinfo{person}{Jimmy Ba}.} \bibinfo{year}{2014}\natexlab{}.
\newblock \showarticletitle{Adam: A method for stochastic optimization}.
\newblock \bibinfo{journal}{\emph{arXiv preprint arXiv:1412.6980}}
  (\bibinfo{year}{2014}).
\newblock


\bibitem[\protect\citeauthoryear{Kitaev, Kaiser, and Levskaya}{Kitaev
  et~al\mbox{.}}{2020}]%
        {reformer}
\bibfield{author}{\bibinfo{person}{Nikita Kitaev}, \bibinfo{person}{{\L}ukasz
  Kaiser}, {and} \bibinfo{person}{Anselm Levskaya}.}
  \bibinfo{year}{2020}\natexlab{}.
\newblock \showarticletitle{Reformer: The efficient transformer}.
\newblock \bibinfo{journal}{\emph{arXiv preprint arXiv:2001.04451}}
  (\bibinfo{year}{2020}).
\newblock


\bibitem[\protect\citeauthoryear{Koren}{Koren}{2009}]%
        {Koren}
\bibfield{author}{\bibinfo{person}{Yehuda Koren}.}
  \bibinfo{year}{2009}\natexlab{}.
\newblock \showarticletitle{Collaborative filtering with temporal dynamics}. In
  \bibinfo{booktitle}{\emph{Proceedings of the 15th ACM SIGKDD international
  conference on Knowledge discovery and data mining}}.
  \bibinfo{pages}{447--456}.
\newblock


\bibitem[\protect\citeauthoryear{Kumar, Irsoy, Ondruska, Iyyer, Bradbury,
  Gulrajani, Zhong, Paulus, and Socher}{Kumar et~al\mbox{.}}{2016}]%
        {kumar}
\bibfield{author}{\bibinfo{person}{Ankit Kumar}, \bibinfo{person}{Ozan Irsoy},
  \bibinfo{person}{Peter Ondruska}, \bibinfo{person}{Mohit Iyyer},
  \bibinfo{person}{James Bradbury}, \bibinfo{person}{Ishaan Gulrajani},
  \bibinfo{person}{Victor Zhong}, \bibinfo{person}{Romain Paulus}, {and}
  \bibinfo{person}{Richard Socher}.} \bibinfo{year}{2016}\natexlab{}.
\newblock \showarticletitle{Ask me anything: Dynamic memory networks for
  natural language processing}. In \bibinfo{booktitle}{\emph{International
  conference on machine learning}}. PMLR, \bibinfo{pages}{1378--1387}.
\newblock


\bibitem[\protect\citeauthoryear{Li, Meng, Zhou, Han, Wu, and Li}{Li
  et~al\mbox{.}}{2020}]%
        {sac}
\bibfield{author}{\bibinfo{person}{Xiaoya Li}, \bibinfo{person}{Yuxian Meng},
  \bibinfo{person}{Mingxin Zhou}, \bibinfo{person}{Qinghong Han},
  \bibinfo{person}{Fei Wu}, {and} \bibinfo{person}{Jiwei Li}.}
  \bibinfo{year}{2020}\natexlab{}.
\newblock \showarticletitle{Sac: Accelerating and structuring self-attention
  via sparse adaptive connection}.
\newblock \bibinfo{journal}{\emph{Advances in Neural Information Processing
  Systems}}  \bibinfo{volume}{33} (\bibinfo{year}{2020}),
  \bibinfo{pages}{16997--17008}.
\newblock


\bibitem[\protect\citeauthoryear{Lian, Zhou, Zhang, Chen, Xie, and Sun}{Lian
  et~al\mbox{.}}{2018}]%
        {lian2018xdeepfm}
\bibfield{author}{\bibinfo{person}{Jianxun Lian}, \bibinfo{person}{Xiaohuan
  Zhou}, \bibinfo{person}{Fuzheng Zhang}, \bibinfo{person}{Zhongxia Chen},
  \bibinfo{person}{Xing Xie}, {and} \bibinfo{person}{Guangzhong Sun}.}
  \bibinfo{year}{2018}\natexlab{}.
\newblock \showarticletitle{xdeepfm: Combining explicit and implicit feature
  interactions for recommender systems}. In
  \bibinfo{booktitle}{\emph{Proceedings of the 24th ACM SIGKDD international
  conference on knowledge discovery \& data mining}}.
  \bibinfo{pages}{1754--1763}.
\newblock


\bibitem[\protect\citeauthoryear{Lin, Feng, Santos, Yu, Xiang, Zhou, and
  Bengio}{Lin et~al\mbox{.}}{2017}]%
        {lin}
\bibfield{author}{\bibinfo{person}{Zhouhan Lin}, \bibinfo{person}{Minwei Feng},
  \bibinfo{person}{Cicero Nogueira~dos Santos}, \bibinfo{person}{Mo Yu},
  \bibinfo{person}{Bing Xiang}, \bibinfo{person}{Bowen Zhou}, {and}
  \bibinfo{person}{Yoshua Bengio}.} \bibinfo{year}{2017}\natexlab{}.
\newblock \showarticletitle{A structured self-attentive sentence embedding}.
\newblock \bibinfo{journal}{\emph{arXiv preprint arXiv:1703.03130}}
  (\bibinfo{year}{2017}).
\newblock


\bibitem[\protect\citeauthoryear{Ma, Ma, Zhang, Sun, Liu, and Coates}{Ma
  et~al\mbox{.}}{2020}]%
        {magnn}
\bibfield{author}{\bibinfo{person}{Chen Ma}, \bibinfo{person}{Liheng Ma},
  \bibinfo{person}{Yingxue Zhang}, \bibinfo{person}{Jianing Sun},
  \bibinfo{person}{Xue Liu}, {and} \bibinfo{person}{Mark Coates}.}
  \bibinfo{year}{2020}\natexlab{}.
\newblock \showarticletitle{Memory augmented graph neural networks for
  sequential recommendation}. In \bibinfo{booktitle}{\emph{Proceedings of the
  AAAI conference on artificial intelligence}}, Vol.~\bibinfo{volume}{34}.
  \bibinfo{pages}{5045--5052}.
\newblock


\bibitem[\protect\citeauthoryear{Matam and Kothapalli}{Matam and
  Kothapalli}{2012}]%
        {gpu2}
\bibfield{author}{\bibinfo{person}{Siva Rama Krishna~Bharadwaj Matam,
  Kiran~Kumar} {and} \bibinfo{person}{Kishore Kothapalli}.}
  \bibinfo{year}{2012}\natexlab{}.
\newblock \showarticletitle{Sparse matrix matrix multiplication on hybrid CPU+
  GPU platforms}. In \bibinfo{booktitle}{\emph{PARCO}}.
\newblock


\bibitem[\protect\citeauthoryear{McAuley, Targett, Shi, and Van
  Den~Hengel}{McAuley et~al\mbox{.}}{2015}]%
        {McAuley}
\bibfield{author}{\bibinfo{person}{Julian McAuley},
  \bibinfo{person}{Christopher Targett}, \bibinfo{person}{Qinfeng Shi}, {and}
  \bibinfo{person}{Anton Van Den~Hengel}.} \bibinfo{year}{2015}\natexlab{}.
\newblock \showarticletitle{Image-based recommendations on styles and
  substitutes}. In \bibinfo{booktitle}{\emph{Proceedings of the 38th
  international ACM SIGIR conference on research and development in information
  retrieval}}. \bibinfo{pages}{43--52}.
\newblock


\bibitem[\protect\citeauthoryear{Pi, Bian, Zhou, Zhu, and Gai}{Pi
  et~al\mbox{.}}{2019}]%
        {MIMN}
\bibfield{author}{\bibinfo{person}{Qi Pi}, \bibinfo{person}{Weijie Bian},
  \bibinfo{person}{Guorui Zhou}, \bibinfo{person}{Xiaoqiang Zhu}, {and}
  \bibinfo{person}{Kun Gai}.} \bibinfo{year}{2019}\natexlab{}.
\newblock \showarticletitle{Practice on long sequential user behavior modeling
  for click-through rate prediction}. In \bibinfo{booktitle}{\emph{Proceedings
  of the 25th ACM SIGKDD International Conference on Knowledge Discovery \&
  Data Mining}}. \bibinfo{pages}{2671--2679}.
\newblock


\bibitem[\protect\citeauthoryear{Qin, Zhang, Wu, Jin, Fang, and Yu}{Qin
  et~al\mbox{.}}{2020}]%
        {ubr4ctr}
\bibfield{author}{\bibinfo{person}{Jiarui Qin}, \bibinfo{person}{Weinan Zhang},
  \bibinfo{person}{Xin Wu}, \bibinfo{person}{Jiarui Jin},
  \bibinfo{person}{Yuchen Fang}, {and} \bibinfo{person}{Yong Yu}.}
  \bibinfo{year}{2020}\natexlab{}.
\newblock \showarticletitle{User behavior retrieval for click-through rate
  prediction}. In \bibinfo{booktitle}{\emph{Proceedings of the 43rd
  International ACM SIGIR Conference on Research and Development in Information
  Retrieval}}. \bibinfo{pages}{2347--2356}.
\newblock


\bibitem[\protect\citeauthoryear{Qin, Sun, Deng, Li, Wei, Lv, Yan, Kong, and
  Zhong}{Qin et~al\mbox{.}}{2022}]%
        {qin2022cosformer}
\bibfield{author}{\bibinfo{person}{Zhen Qin}, \bibinfo{person}{Weixuan Sun},
  \bibinfo{person}{Hui Deng}, \bibinfo{person}{Dongxu Li},
  \bibinfo{person}{Yunshen Wei}, \bibinfo{person}{Baohong Lv},
  \bibinfo{person}{Junjie Yan}, \bibinfo{person}{Lingpeng Kong}, {and}
  \bibinfo{person}{Yiran Zhong}.} \bibinfo{year}{2022}\natexlab{}.
\newblock \showarticletitle{cosFormer: Rethinking Softmax in Attention}.
\newblock \bibinfo{journal}{\emph{arXiv preprint arXiv:2202.08791}}
  (\bibinfo{year}{2022}).
\newblock


\bibitem[\protect\citeauthoryear{Rae, Potapenko, Jayakumar, and Lillicrap}{Rae
  et~al\mbox{.}}{2019}]%
        {rae2019compressive}
\bibfield{author}{\bibinfo{person}{Jack~W Rae}, \bibinfo{person}{Anna
  Potapenko}, \bibinfo{person}{Siddhant~M Jayakumar}, {and}
  \bibinfo{person}{Timothy~P Lillicrap}.} \bibinfo{year}{2019}\natexlab{}.
\newblock \showarticletitle{Compressive transformers for long-range sequence
  modelling}.
\newblock \bibinfo{journal}{\emph{arXiv preprint arXiv:1911.05507}}
  (\bibinfo{year}{2019}).
\newblock


\bibitem[\protect\citeauthoryear{Ramachandran and Sohmshetty}{Ramachandran and
  Sohmshetty}{2017}]%
        {sachithanandam}
\bibfield{author}{\bibinfo{person}{Govardana~Sachithanandam Ramachandran} {and}
  \bibinfo{person}{Ajay Sohmshetty}.} \bibinfo{year}{2017}\natexlab{}.
\newblock \showarticletitle{Ask me even more: dynamic memory tensor networks
  (extended model)}.
\newblock \bibinfo{journal}{\emph{arXiv preprint arXiv:1703.03939}}
  (\bibinfo{year}{2017}).
\newblock


\bibitem[\protect\citeauthoryear{Ramachandran, Parmar, Vaswani, Bello,
  Levskaya, and Shlens}{Ramachandran et~al\mbox{.}}{2019}]%
        {parmar}
\bibfield{author}{\bibinfo{person}{Prajit Ramachandran}, \bibinfo{person}{Niki
  Parmar}, \bibinfo{person}{Ashish Vaswani}, \bibinfo{person}{Irwan Bello},
  \bibinfo{person}{Anselm Levskaya}, {and} \bibinfo{person}{Jon Shlens}.}
  \bibinfo{year}{2019}\natexlab{}.
\newblock \showarticletitle{Stand-alone self-attention in vision models}.
\newblock \bibinfo{journal}{\emph{Advances in Neural Information Processing
  Systems}}  \bibinfo{volume}{32} (\bibinfo{year}{2019}).
\newblock


\bibitem[\protect\citeauthoryear{Sodhani, Chandar, and Bengio}{Sodhani
  et~al\mbox{.}}{2020}]%
        {sodhani2020toward}
\bibfield{author}{\bibinfo{person}{Shagun Sodhani}, \bibinfo{person}{Sarath
  Chandar}, {and} \bibinfo{person}{Yoshua Bengio}.}
  \bibinfo{year}{2020}\natexlab{}.
\newblock \showarticletitle{Toward training recurrent neural networks for
  lifelong learning}.
\newblock \bibinfo{journal}{\emph{Neural computation}} \bibinfo{volume}{32},
  \bibinfo{number}{1} (\bibinfo{year}{2020}), \bibinfo{pages}{1--35}.
\newblock


\bibitem[\protect\citeauthoryear{Tan, Zhang, Liu, Huang, Yang, Zhou, Hu,
  et~al\mbox{.}}{Tan et~al\mbox{.}}{2021}]%
        {tan2021dynamic}
\bibfield{author}{\bibinfo{person}{Qiaoyu Tan}, \bibinfo{person}{Jianwei
  Zhang}, \bibinfo{person}{Ninghao Liu}, \bibinfo{person}{Xiao Huang},
  \bibinfo{person}{Hongxia Yang}, \bibinfo{person}{Jignren Zhou},
  \bibinfo{person}{Xia Hu}, {et~al\mbox{.}}} \bibinfo{year}{2021}\natexlab{}.
\newblock \showarticletitle{Dynamic memory based attention network for
  sequential recommendation}.
\newblock \bibinfo{journal}{\emph{arXiv preprint arXiv:2102.09269}}
  (\bibinfo{year}{2021}).
\newblock


\bibitem[\protect\citeauthoryear{Vaswani, Shazeer, Parmar, Uszkoreit, Jones,
  Gomez, Kaiser, and Polosukhin}{Vaswani et~al\mbox{.}}{2017}]%
        {transformer}
\bibfield{author}{\bibinfo{person}{Ashish Vaswani}, \bibinfo{person}{Noam
  Shazeer}, \bibinfo{person}{Niki Parmar}, \bibinfo{person}{Jakob Uszkoreit},
  \bibinfo{person}{Llion Jones}, \bibinfo{person}{Aidan~N Gomez},
  \bibinfo{person}{{\L}ukasz Kaiser}, {and} \bibinfo{person}{Illia
  Polosukhin}.} \bibinfo{year}{2017}\natexlab{}.
\newblock \showarticletitle{Attention is all you need}.
\newblock \bibinfo{journal}{\emph{Advances in neural information processing
  systems}}  \bibinfo{volume}{30} (\bibinfo{year}{2017}).
\newblock


\bibitem[\protect\citeauthoryear{Vig and Belinkov}{Vig and Belinkov}{2019}]%
        {vig2019analyzing}
\bibfield{author}{\bibinfo{person}{Jesse Vig} {and} \bibinfo{person}{Yonatan
  Belinkov}.} \bibinfo{year}{2019}\natexlab{}.
\newblock \showarticletitle{Analyzing the structure of attention in a
  transformer language model}.
\newblock \bibinfo{journal}{\emph{arXiv preprint arXiv:1906.04284}}
  (\bibinfo{year}{2019}).
\newblock


\bibitem[\protect\citeauthoryear{Wang, Li, Khabsa, Fang, and Ma}{Wang
  et~al\mbox{.}}{2020}]%
        {linformer}
\bibfield{author}{\bibinfo{person}{Sinong Wang}, \bibinfo{person}{Belinda~Z
  Li}, \bibinfo{person}{Madian Khabsa}, \bibinfo{person}{Han Fang}, {and}
  \bibinfo{person}{Hao Ma}.} \bibinfo{year}{2020}\natexlab{}.
\newblock \showarticletitle{Linformer: Self-attention with linear complexity}.
\newblock \bibinfo{journal}{\emph{arXiv preprint arXiv:2006.04768}}
  (\bibinfo{year}{2020}).
\newblock


\bibitem[\protect\citeauthoryear{Weston, Chopra, and Bordes}{Weston
  et~al\mbox{.}}{2014}]%
        {weston}
\bibfield{author}{\bibinfo{person}{Jason Weston}, \bibinfo{person}{Sumit
  Chopra}, {and} \bibinfo{person}{Antoine Bordes}.}
  \bibinfo{year}{2014}\natexlab{}.
\newblock \showarticletitle{Memory Networks}. In
  \bibinfo{booktitle}{\emph{arXiv preprint arXiv:1410.3916}}.
\newblock


\bibitem[\protect\citeauthoryear{Wu, Ren, Yu, Chen, Zhang, and Zhu}{Wu
  et~al\mbox{.}}{2016}]%
        {wu}
\bibfield{author}{\bibinfo{person}{Sai Wu}, \bibinfo{person}{Weichao Ren},
  \bibinfo{person}{Chengchao Yu}, \bibinfo{person}{Gang Chen},
  \bibinfo{person}{Dongxiang Zhang}, {and} \bibinfo{person}{Jingbo Zhu}.}
  \bibinfo{year}{2016}\natexlab{}.
\newblock \showarticletitle{Personal recommendation using deep recurrent neural
  networks in NetEase}. In \bibinfo{booktitle}{\emph{2016 IEEE 32nd
  international conference on data engineering (ICDE)}}. IEEE,
  \bibinfo{pages}{1218--1229}.
\newblock


\bibitem[\protect\citeauthoryear{Wu, Tang, Zhu, Wang, Xie, and Tan}{Wu
  et~al\mbox{.}}{2019}]%
        {gnn}
\bibfield{author}{\bibinfo{person}{Shu Wu}, \bibinfo{person}{Yuyuan Tang},
  \bibinfo{person}{Yanqiao Zhu}, \bibinfo{person}{Liang Wang},
  \bibinfo{person}{Xing Xie}, {and} \bibinfo{person}{Tieniu Tan}.}
  \bibinfo{year}{2019}\natexlab{}.
\newblock \showarticletitle{Session-based recommendation with graph neural
  networks}. In \bibinfo{booktitle}{\emph{Proceedings of the AAAI conference on
  artificial intelligence}}, Vol.~\bibinfo{volume}{33}.
  \bibinfo{pages}{346--353}.
\newblock


\bibitem[\protect\citeauthoryear{Xiong, Merity, and Socher}{Xiong
  et~al\mbox{.}}{2016}]%
        {xiong}
\bibfield{author}{\bibinfo{person}{Caiming Xiong}, \bibinfo{person}{Stephen
  Merity}, {and} \bibinfo{person}{Richard Socher}.}
  \bibinfo{year}{2016}\natexlab{}.
\newblock \showarticletitle{Dynamic memory networks for visual and textual
  question answering}. In \bibinfo{booktitle}{\emph{International conference on
  machine learning}}. PMLR, \bibinfo{pages}{2397--2406}.
\newblock


\bibitem[\protect\citeauthoryear{Xu, Ba, Kiros, Cho, Courville, Salakhudinov,
  Zemel, and Bengio}{Xu et~al\mbox{.}}{2015}]%
        {xu}
\bibfield{author}{\bibinfo{person}{Kelvin Xu}, \bibinfo{person}{Jimmy Ba},
  \bibinfo{person}{Ryan Kiros}, \bibinfo{person}{Kyunghyun Cho},
  \bibinfo{person}{Aaron Courville}, \bibinfo{person}{Ruslan Salakhudinov},
  \bibinfo{person}{Rich Zemel}, {and} \bibinfo{person}{Yoshua Bengio}.}
  \bibinfo{year}{2015}\natexlab{}.
\newblock \showarticletitle{Show, attend and tell: Neural image caption
  generation with visual attention}. In \bibinfo{booktitle}{\emph{International
  conference on machine learning}}. PMLR, \bibinfo{pages}{2048--2057}.
\newblock


\bibitem[\protect\citeauthoryear{Yu, Lian, Mahmoody, Liu, and Xie}{Yu
  et~al\mbox{.}}{2019}]%
        {yu2019adaptive}
\bibfield{author}{\bibinfo{person}{Zeping Yu}, \bibinfo{person}{Jianxun Lian},
  \bibinfo{person}{Ahmad Mahmoody}, \bibinfo{person}{Gongshen Liu}, {and}
  \bibinfo{person}{Xing Xie}.} \bibinfo{year}{2019}\natexlab{}.
\newblock \showarticletitle{Adaptive User Modeling with Long and Short-Term
  Preferences for Personalized Recommendation.}. In
  \bibinfo{booktitle}{\emph{IJCAI}}. \bibinfo{pages}{4213--4219}.
\newblock


\bibitem[\protect\citeauthoryear{Yuan, Karatzoglou, Arapakis, Jose, and
  He}{Yuan et~al\mbox{.}}{2019}]%
        {yuan2019simple}
\bibfield{author}{\bibinfo{person}{Fajie Yuan}, \bibinfo{person}{Alexandros
  Karatzoglou}, \bibinfo{person}{Ioannis Arapakis}, \bibinfo{person}{Joemon~M
  Jose}, {and} \bibinfo{person}{Xiangnan He}.} \bibinfo{year}{2019}\natexlab{}.
\newblock \showarticletitle{A simple convolutional generative network for next
  item recommendation}. In \bibinfo{booktitle}{\emph{Proceedings of the Twelfth
  ACM International Conference on Web Search and Data Mining}}.
  \bibinfo{pages}{582--590}.
\newblock


\bibitem[\protect\citeauthoryear{Zhou, Bai, Song, Liu, Zhao, Chen, and
  Gao}{Zhou et~al\mbox{.}}{2018a}]%
        {atrank}
\bibfield{author}{\bibinfo{person}{Chang Zhou}, \bibinfo{person}{Jinze Bai},
  \bibinfo{person}{Junshuai Song}, \bibinfo{person}{Xiaofei Liu},
  \bibinfo{person}{Zhengchao Zhao}, \bibinfo{person}{Xiusi Chen}, {and}
  \bibinfo{person}{Jun Gao}.} \bibinfo{year}{2018}\natexlab{a}.
\newblock \showarticletitle{Atrank: An attention-based user behavior modeling
  framework for recommendation}. In \bibinfo{booktitle}{\emph{Thirty-Second
  AAAI Conference on Artificial Intelligence}}.
\newblock


\bibitem[\protect\citeauthoryear{Zhou, Mou, Fan, Pi, Bian, Zhou, Zhu, and
  Gai}{Zhou et~al\mbox{.}}{2019}]%
        {dien}
\bibfield{author}{\bibinfo{person}{Guorui Zhou}, \bibinfo{person}{Na Mou},
  \bibinfo{person}{Ying Fan}, \bibinfo{person}{Qi Pi}, \bibinfo{person}{Weijie
  Bian}, \bibinfo{person}{Chang Zhou}, \bibinfo{person}{Xiaoqiang Zhu}, {and}
  \bibinfo{person}{Kun Gai}.} \bibinfo{year}{2019}\natexlab{}.
\newblock \showarticletitle{Deep interest evolution network for click-through
  rate prediction}. In \bibinfo{booktitle}{\emph{Proceedings of the AAAI
  conference on artificial intelligence}}, Vol.~\bibinfo{volume}{33}.
  \bibinfo{pages}{5941--5948}.
\newblock


\bibitem[\protect\citeauthoryear{Zhou, Zhu, Song, Fan, Zhu, Ma, Yan, Jin, Li,
  and Gai}{Zhou et~al\mbox{.}}{2018b}]%
        {din}
\bibfield{author}{\bibinfo{person}{Guorui Zhou}, \bibinfo{person}{Xiaoqiang
  Zhu}, \bibinfo{person}{Chenru Song}, \bibinfo{person}{Ying Fan},
  \bibinfo{person}{Han Zhu}, \bibinfo{person}{Xiao Ma},
  \bibinfo{person}{Yanghui Yan}, \bibinfo{person}{Junqi Jin},
  \bibinfo{person}{Han Li}, {and} \bibinfo{person}{Kun Gai}.}
  \bibinfo{year}{2018}\natexlab{b}.
\newblock \showarticletitle{Deep interest network for click-through rate
  prediction}. In \bibinfo{booktitle}{\emph{Proceedings of the 24th ACM SIGKDD
  international conference on knowledge discovery \& data mining}}.
  \bibinfo{pages}{1059--1068}.
\newblock


\bibitem[\protect\citeauthoryear{Zhou, Zhang, Peng, Zhang, Li, Xiong, and
  Zhang}{Zhou et~al\mbox{.}}{2021}]%
        {informer}
\bibfield{author}{\bibinfo{person}{Haoyi Zhou}, \bibinfo{person}{Shanghang
  Zhang}, \bibinfo{person}{Jieqi Peng}, \bibinfo{person}{Shuai Zhang},
  \bibinfo{person}{Jianxin Li}, \bibinfo{person}{Hui Xiong}, {and}
  \bibinfo{person}{Wancai Zhang}.} \bibinfo{year}{2021}\natexlab{}.
\newblock \showarticletitle{Informer: Beyond efficient transformer for long
  sequence time-series forecasting}. In \bibinfo{booktitle}{\emph{Proceedings
  of AAAI}}.
\newblock


\end{thebibliography}

\appendix

\end{document}